\documentclass[%
 aip,
 amsmath,amssymb,
 reprint,%
]{revtex4-1}

\usepackage{graphicx}% Include figure files
\usepackage{dcolumn}% Align table columns on decimal point
\usepackage{bm}% bold math
\usepackage[utf8]{inputenc}
\usepackage[T1]{fontenc}
\usepackage{mathptmx}
\begin{document}
\preprint{AIP/123-QED}

\title{Experimental demonstration of tunable graphene-polaritonic hyperbolic metamaterial}
% Force line breaks with \\

\author{Jeremy Brouillet}
 \affiliation{Thomas J. Watson Laboratories of Applied Physics, California Institute of Technology, California 91125, USA}%Lines break automatically or can be forced with \\
\author{Georgia T. Papadakis}
 \altaffiliation[Present address:]{ Department of Electrical Engineering, Ginzton Laboratory, Stanford University, California 94305, USA;
gpapadak@stanford.edu}
 \email{gpapadak@stanford.edu}
 \affiliation{Thomas J. Watson Laboratories of Applied Physics, California Institute of Technology, California 91125, USA}%Lines  
\author{Harry A. Atwater}%
\affiliation{Thomas J. Watson Laboratories of Applied Physics, California Institute of Technology, California 91125, USA}

%Authors' institution and/or address%\\This line break forced with \textbackslash\textbackslash
%}%

\date{\today}% It is always \today, today,

\begin{abstract}
Tuning the macroscopic dielectric response on demand holds potential for actively tunable metaphotonics and optical devices. In recent years, graphene has been extensively investigated as a tunable element in nanophotonics. Significant theoretical work has been devoted on the tuning the hyperbolic properties of graphene/dielectric heterostructures, however, until now, such a motif has not been demonstrated experimentally. Here we focus on a graphene/polaritonic dielectric metamaterial, with strong optical resonances arising from the polar response of the dielectric, which are, in general, difficult to actively control. By controlling the doping level of graphene via external bias we experimentally demonstrate a wide range of tunability from a Fermi level of $E_\mathrm{F}=0$ eV to $E_\mathrm{F}=0.5$ eV, which yields an effective epsilon-near-zero crossing and tunable dielectric properties, verified through spectroscopic ellipsometry and transmission measurements.
\end{abstract}

\maketitle

% \section{\label{sec:level1}Introduction }

%%%%%%%%%%%%%%%%%%%%%%%%%%%%%%%%%%%%%%% Georgia Old intro
\par{Spectral tunability is key for controlling light-matter interactions, critical for many applications including emission control, surface enhanced spectroscopy, sensing, and thermal control. Particularly in the subwavelength range, tuning plasmonic resonances has been essential in controlling color, typically achieved by controlling the size of plasmonic nanoparticles, antennas and metamaterials \cite{lu2014enhancing,zhou2014experiment,jacob2010engineering,schuller2010plasmonics}. In obtaining a large range of spectral tunability, it is preferable to operate near an optical resonance rather than a broadband plasmonic response. Nevertheless, it is in general easier to tune a broadband optical response rather than a resonant one since resonances in nanophotonics typically entail subwavelength-scale geometrical features.}

\par{From a very wide range of recently investigated metamaterials and heterostructures for spectral control, particular emphasis has been given to hyperbolic media, due to enhanced light-matter interactions arising from a larger range of wavenumbers available for propagating modes \cite{smith2000composite}. These media are in generally uniaxial and support a hyperbolic frequency dispersion given by the equation \cite{jacob2010engineering,poddubny2013hyperbolic,smith2003electromagnetic,papadakis2015retrieval}
\begin{equation}\label{eq:1}
    \frac{k_{x}^{2}+k_{y}^{2}}{\epsilon_\mathrm{e}}+\frac{k_{z}^{2}}{\epsilon_\mathrm{o}}=\frac{\omega^2}{c^{2}}
\end{equation}
where $\epsilon_{o}$ and $\epsilon_{e}$ refer to the ordinary (in-plane) and extraordinary (out-of-plane) dielectric permittivity, respectively. Due to the different sign in $\epsilon_{o}$ and $\epsilon_{e}$, upon fixing the frequency $\omega$, the isofrequency diagram of the relevant electromagnetic modes opens up into a hyperbola, giving rise to a very large density of optical states, promising for waveguiding \cite{Babicheva_WG}, emission engineering and Purcell enhancement \cite{cortes2019ground,lu2014enhancing,zhou2014experiment} thermal photonics \cite{Biehs_BB}, lasing \cite{fang2010lasing}, and imaging \cite{Zhang_Science,weile2007electromagnetic}. Particularly, near the epsilon-near-zero frequency crossing of either $\epsilon_\mathrm{o}$ or $\epsilon_\mathrm{e}$, many exciting phenomena can be supported, the most prominent of which is light propagation with near-zero phase advance \cite{maas2013experimental,mahmoud2014wave,engheta2013pursuing}.}

\par{There has been significant effort in frequency-tuning of the optical response of hyperbolic metamaterials \cite{papadakis_tunable,poddubny2013hyperbolic,FanJon_HMM,Tunable_HMM_review}. For this, particular interest holds the case of graphene, a well-studied monolayer material for electronics \cite{novoselov2005two} and in infrared photonics \cite{andersen2010graphene}. Namely, the dielectric properties of graphene can be dynamically tuned via optical pumping \cite{ryzhii2007negative}, or with electrostatic modulation of its carrier concentration with field-effect gating \cite{vakil2011transformation,polini2008plasmons}, often targeting tunable plasmonic properties \cite{hwang2007dielectric,brar2013highly}. The high degree of localization of graphene plasmons, together with the dielectric tunability of graphene provides a promising platform for investigating tunable graphene-based hyperbolic metamaterials. There has already been considerable theoretical effort in the past decade to understand the properties of tunable graphene metamaterials \cite{linder2016graphene,andryieuski2013graphene,xiong2018ultra,zainud2017frequency,al2014control}, with significant focus on the potential of tuning hyperbolic properties of graphene/dielectric planar heterostructures \cite{poddubny2013hyperbolic,iorsh2013hyperbolic,othman2013graphene,Janaszek2016}. There have previously been experimental demonstrations of graphene-based hyperbolic media \cite{chang2016realization,Basov_GHMM}, nevertheless, the reported properties have remained fixed at the time of fabrication. No post-fabrication way to control the dielectric permittivity tensor ($\epsilon_\mathrm{o}$ and $\epsilon_\mathrm{e}$ in Eq. \ref{eq:1}) has been reported until now.}

\par{Gating graphene when integrated with dielectric layers is difficult due to graphene's two-dimensional nature with weak out-of-plane Van der Waals bonds that yield poor adhesion to most dielectric substrates. Furthermore, large-area graphene sheets on the order of mm$^2$’s with gate-induced tunability are needed to perform metamaterial optical measurements at infrared frequencies. Exfoliated flakes are generally limited to sizes of 10s of $\mu$m, so large-area graphene samples grown by chemical vapor deposition and subsequently transferred from their growth substrates, are necessary. Additionally, deposition of large-area thin dielectric layers on graphene is challenging. Films prepared by electron-beam evaporation exhibit thermal stress-induced delamination \cite{mcnerny2014direct}. Films grown by atomic layer deposition (ALD) with an H$_{2}$O precursors exhibit difficulty in bonding to chemically-inert hydrophobic graphene \cite{park2014wetting}, whereas ozone-based ALD processes oxidize graphene.}

\par{Here, we discuss how we overcome these challenges and are, thus, able to tune a graphene-based hyperbolic metamaterial unit cell for a wide range of doping levels in graphene translating to a Fermi level that ranges from $E_\mathrm{F}=0$ eV to $E_\mathrm{F}=0.5$ eV, without dielectric breakdown. Previous theoretical proposals have considered non-dispersive dielectric materials \cite{poddubny2013hyperbolic,iorsh2013hyperbolic,othman2013graphene,Janaszek2016}, thereby yielding a broadband hyperbolic response. By contrast, here, we consider a polaritonic dielectric material, namely SiO$_\mathrm{2}$. The polaritonic resonances that all polar materials exhibit at infrared frequencies, at their Reststrahlen band, are typically not tunable, as they constitute a fundamental material property. Nevertheless, we show here that, upon the integration of graphene, it is feasible to actively tune these polaritonic resonances. Graphene provides a tunable character to the in-plane response of the composite graphene/SiO$_\mathrm{2}$ heterostructure, and its plasmonic nature assigns a hyperbolic frequency region near the polar resonance of SiO$_\mathrm{2}$, at a free-space wavelength of $20$ $\mu$m. We are therefore able to experimentally observe, through multi-angle spectroscopic ellipsometry and transmittance measurements, a tunable epsilon-near zero permittivity along the in-plane direction near the surface phonon polaritonic resonance while leaving the out-of-plane response unchanged (due to the two-dimensional nature of graphene), thereby yielding a widely tunable hyperbolic response.}

\begin{figure}
\centering
\includegraphics[width=0.9\columnwidth]{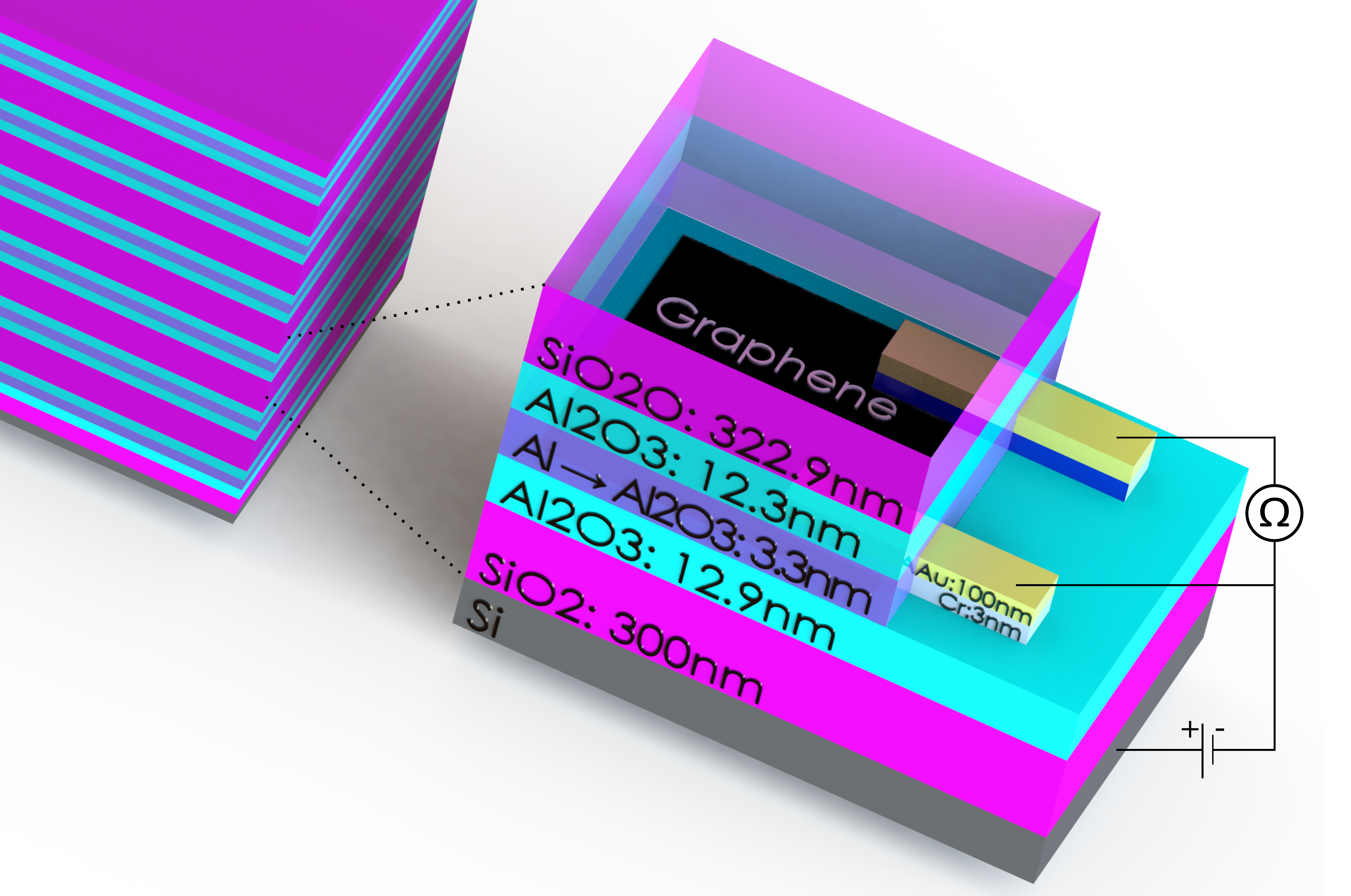}% 
\caption{\label{fig:diagramFig} Left: Schematic of a theoretical metamaterial stack. Right: Schematic of the fabricated individual device. The layers: Lightly-doped silicon substrate, thermally-grown SiO2, Al$_{2}$O$_{3}$, transferred chemical-vapor deposited (CVD) graphene, Al$_{2}$O$_{3}$, and plasma-enhance chemical vapor deposition SiO$_{2}$. The thin layers of Al$_{2}$O$_{3}$ are necessary for the feasibility of the fabrication. The thick SiO$_{2}$ contribute to the majority of the dielectric response. Contacts are added to gate and measure the resistance of the graphene. The graphene is tuned by gating against the back silicon substrate.}
\end{figure}

\par{The metamaterial under consideration is depicted in Fig.~\ref{fig:diagramFig}, and is composed of a graphene monolayer sandwitched in between two SiO$_\mathrm{2}$ layers of thickness $300$ nm. The alumina (Al$_\mathrm{2}$O$\mathrm{3}$) layers depicted in Fig.~\ref{fig:diagramFig} have thickness  thickness $0.5$ nm and are placed to prevent poor graphene adhesion. Particularly, a viable dielectric deposition method was developed consisting of functionalization of the surface by deposition of trimethylaluminium (TMA) \cite{lee2010characteristics} or an aluminum nucleation layer \cite{kim2009realization} to create a seed layer for additional deposition. A suitably thin layer of aluminum is needed  so that it can fully oxidize and not compromise the electrical gating of the graphene. We found that deposition of AL$_{2}$O$_{3}$ via plasma-enhanced chemical vapor deposition (PECVD) resulted in reduced thermal stress and avoided delamination. The graphene is grown by chemical vapor deposition (CVD) and transferred onto the thermal oxide, whereas the top SiO$_\mathrm{2}$ film is deposited by plasma-enhanced chemical vapor deposition (PECVD). The thickness of the film layers were measured by both a thin film analyzer and visible ellipsometry with a qualitative agreement of 2nm. Lithographically-defined patterns were used to deposit 3nm/100nm of Cr/Au contacts on the graphene layer, and were used to gate the graphene monolayer against the silicon substrate, which serves as the back-side contact for field-effect tuning.}

\par
\begin{figure}
\centering
\includegraphics[width=0.8\columnwidth]{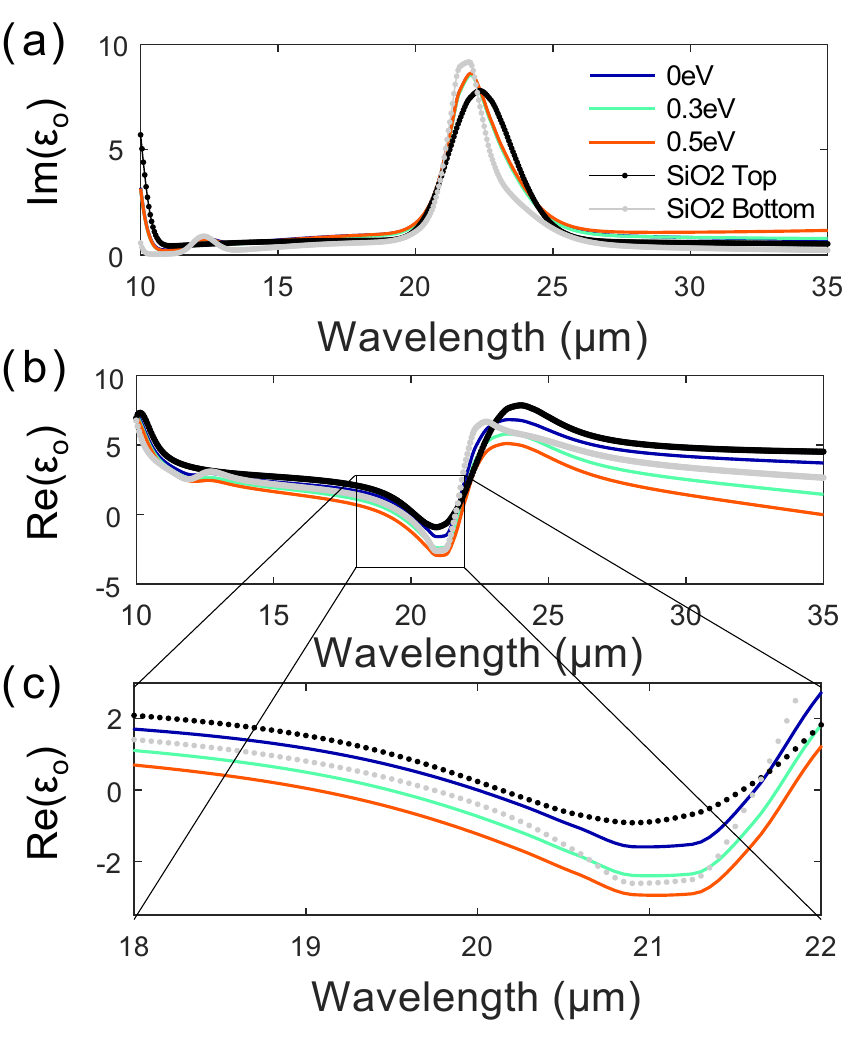}% Here is how to import EPS art
\caption{\label{fig:fig2epsilon} Ellipsometrically derived $\epsilon_\mathrm{o}$ for the graphene/SiO$_\mathrm{2}$ metamaterial of Fig.~\ref{fig:diagramFig}, under applied bias, for three different Fermi levels E$_\mathrm{F}$. Grey and black curves correspond to the homogeneous dielectric permittivity of the bottom and top $_\mathrm{2}$ films, respectively. (a) Imaginary part, and (b) real part. (c) Inset showing the epsilon-near-zero regime of $\epsilon_\mathrm{o}$ at different E$_\mathrm{F}$'s.}
\end{figure}

{Since the composite in Fig.~\ref{fig:diagramFig} is extremely subwavelength to infrared light, one can homogenize it and assign an effective in-plane and out-of-plane dielectric response, namely $\epsilon_\mathrm{o}$ and $\epsilon_\mathrm{e}$ \cite{papadakis2015retrieval}. The two-dimensional nature of graphene leaves the out-of-plane response unaffected, therefore in the out-of-plane direction, this metamaterial behaves to far-field radiation effectively as bulk SiO$_\mathrm{2}$. By striking contrast, by electrostatically tuning the graphene carrier we can shift the epsilon-near-zero point of $\epsilon_\mathrm{o}$, and therefore control the hyperbolicity of the heterostructure as shown in Fig.~\ref{fig:fig2epsilon}.}

\par{In estimating the Fermi level to which we can actively tune the doping level in graphene, we use a capacitor model based on the materials between the gate and the applied voltage\cite{luxmoore2014strong}.
\begin{equation}
    E_{f} = 0.031\sqrt{V-V_\mathrm{Dirac}}.
\end{equation}
Experimentally, the location of the Dirac peak was determined via measuring change in sheet resistance. Furthermore, we use the Kubo formula \cite{falkovsky2008optical} calculate the sheet conductance $\sigma$ from the E$_{f}$ of graphene. This value can be used to compute the transfer matrix for graphene \cite{papadakis2018optical}.
\begin{center}
\[
\overleftrightarrow{G}
=
\begin{bmatrix}
   1 & 0 \\
  4\pi\sigma/c & 0 \\
\end{bmatrix}
\]
\end{center}
We utilize the transfer matrix approach \cite{wiley1988sons}, accounting for graphene via $\overleftrightarrow{G}$, and obtain the complex scattering amplitudes of the fields at different Fermi levels $E_\mathrm{F}$. In these calculations, fabrication and material imperfections are removed by having, a priori, measured experimentally the individual layer thicknesses and optical constants of all thin films in the metamaterial, with ellipsometry. For example, in Fig.~\ref{fig:fig2epsilon}(a) and (b) we show the experimentally determined dielectric permittivity of the top and bottom SiO$_\mathrm{2}$ films shown in Fig.~\ref{fig:diagramFig}, where their small differences are are expected since the top SiO$_\mathrm{2}$ is deposited via PECVD whereas the bottom one is thermally grown. The scattering amplitudes are fed into previously developed parameter retrieval approaches \cite{papadakis2015retrieval}, from which we obtain an effective uniaxial tensorial dielectric permittivity $\epsilon=\mathrm{diag}(\epsilon_\mathrm{o},\epsilon_\mathrm{o},\epsilon_\mathrm{e})$ that characterizes the metamaterial composite. This process is repeated at different gating voltages $V$, in other words for different Fermi levels $E_\mathrm{F}$.}

\par{By taking spectroscopic ellipsometry measurements of the full metamaterial stack of Fig.~\ref{fig:diagramFig}, we perform an ellipsometric fitting where we use the effective dielectric permittivity  $\epsilon=\mathrm{diag}(\epsilon_\mathrm{o},\epsilon_\mathrm{o},\epsilon_\mathrm{e})$ as a model to fit to the experimental data, namely the ellipsometric observables $\Psi$ and $\Delta$. In Fig.~\ref{fig:fig2epsilon}(a) and (b) we show the imaginary and real part of the ellipsometrically-derived in-plane permittivity $\epsilon_\mathrm{o}$, at different Fermi levels $E_\mathrm{F}$. We note that the out-of-plane effective permittivity $\epsilon_\mathrm{e}$ is not tunable as described above, and therefore is omitted. There resonant character of $\epsilon_\mathrm{o}$ near the regime of $20$ $\mu$m is attributed to the surface phonon polaritonic resonance of SiO$_\mathrm{2}$ at this wavelength, nevertheless this resonance has now become tunable via incorporation of a monolayer-thick graphene sheet in between SiO$_\mathrm{2}$ films. As can be clearly seen in \ref{fig:fig2epsilon}(c), by gradually tuning the Fermi level of graphene from $E_\mathrm{F}=0$ eV (blue curves) to $E_\mathrm{F}=0.3$ eV (green curves) to $E_\mathrm{F}=0.5$ eV, we redshift the infrared response of the metamaterial by approximately a micron, i.e. from a near-zero crossing at $20$ $\mu$m under no bias to $19$ $\mu$m under large applied bias. Redshifting is expected as a response of applied bias because the electrostatic doping induces additional charge carriers in the graphene sheet, hence making the composite medium more metallic.}

\par
\begin{figure}[b!]
\centering
\includegraphics[width=0.7\columnwidth]{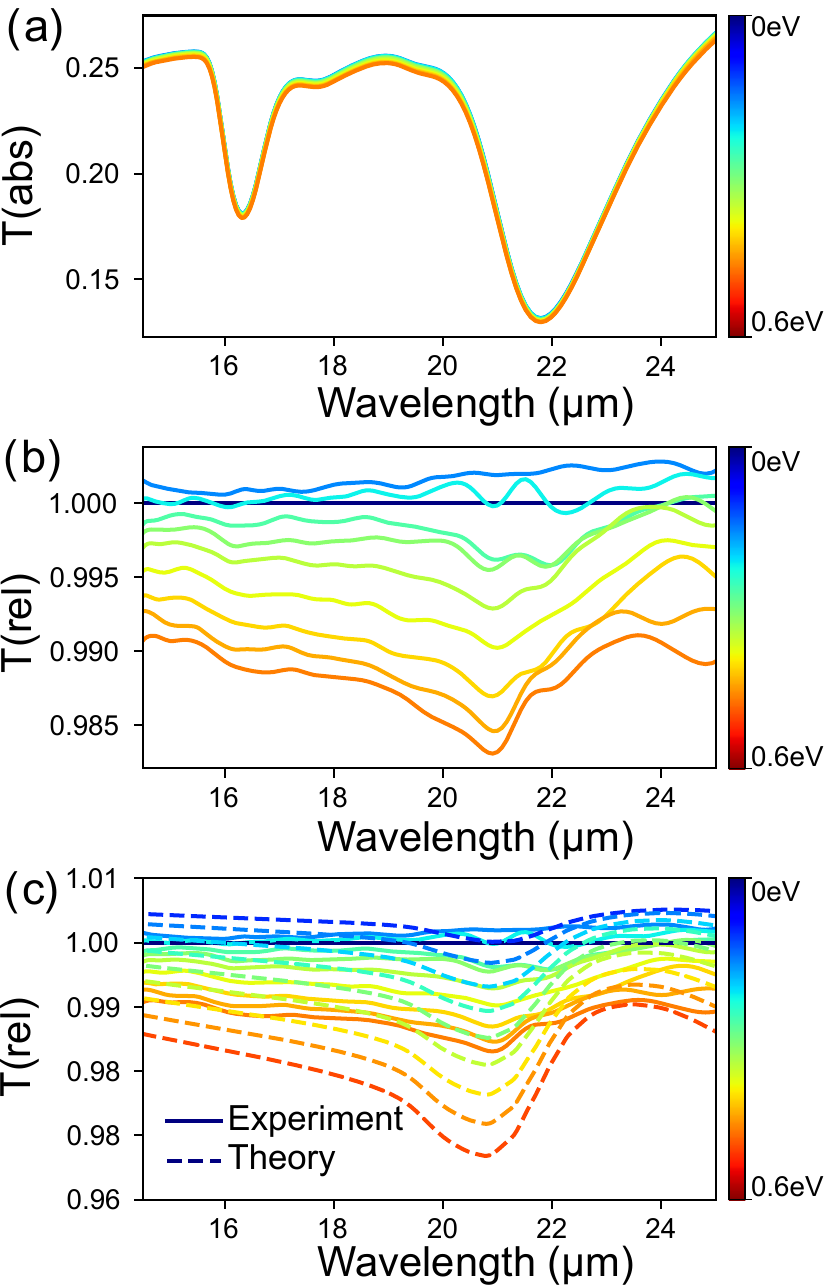}% 
\caption{\label{fig:rawDataFig}  (a) Absolute FTIR transmission over a range of Fermi levels. (b) Experimental data normalized to E$_\mathrm{F}$=0, the Dirac point. (c) Experiment compared with theory, based on ellipsometric fits the thickness and optical properties constituent layers. Normalized to E$_\mathrm{F}$=0. Deviations arise due to hysteresis of the graphene induced by charge trapping.}
\end{figure}

{In addition to spectroscopic ellipsometry, we perform Fourier-transform infrared spectroscopy (FTIR) to measure the sample transmission, and compare with the results of spectroscopic ellipsometry shown above, derived based on initial parameter retrieval-based derivation of $\epsilon=\mathrm{diag}(\epsilon_\mathrm{o},\epsilon_\mathrm{o},\epsilon_\mathrm{e})$. Electrostatically gating the graphene induces changes in the transmission of the composite metamaterial, as shown in Fig. ~\ref{fig:rawDataFig}. Namely, as mentioned above, gating the graphene monolayer makes the composite metamaterial more metallic and, therefore, less transmissive, as shown with the colormap in Figs. ~\ref{fig:rawDataFig}(b) and (c). The dips near the wavelengths of $16$ $\mu$m and $20$ $\mu$m correspond to the two surface phonon polariton resonances of SiO$_\mathrm{2}$, where the material absorbs resonantly, resulting in low transmittance. We note that, experimentally, graphene exhibits hysteresis, which is attributed to defects induced by the deposition of the aluminum layer, resulting in the discrepancies between experiment and theory. As the graphene is tuned, the Dirac peak shifts in the direction of applied bias, causing the sample to experience a reduced E$_{f}$, giving qualitative experimental agreement with theory without fitting parameters as can be seen in Fig. ~\ref{fig:rawDataFig}(c).}

\par
\begin{figure}
\centering
\includegraphics[width=0.7\columnwidth]{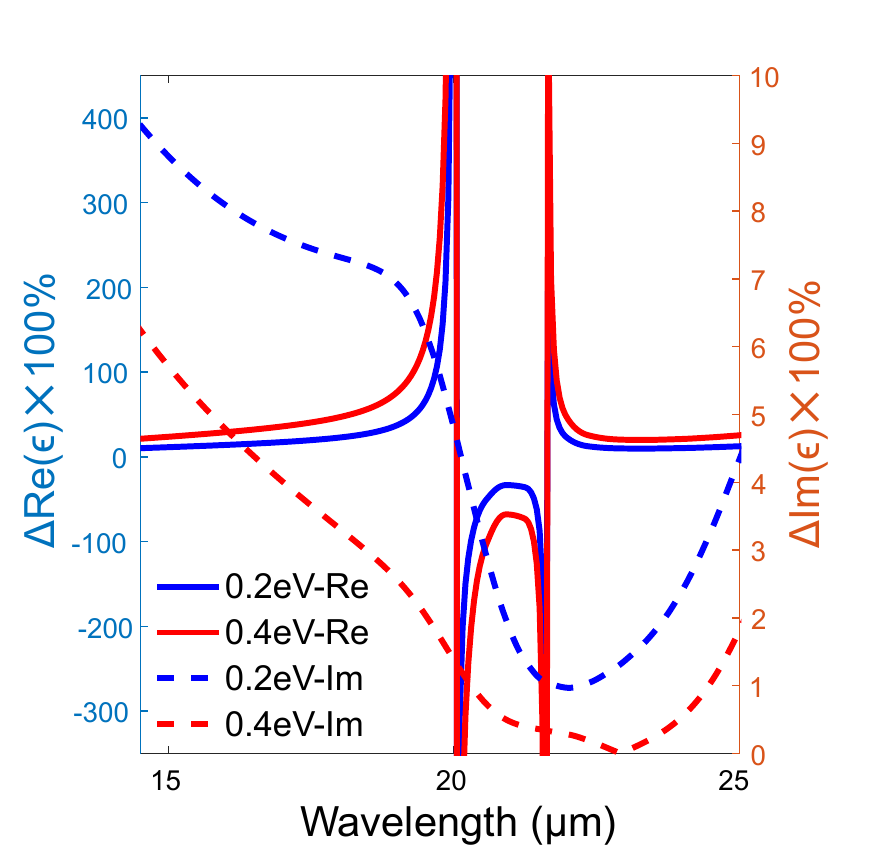}% 
\caption{\label{fig:epsilonFig} Relative change of $\epsilon_\mathrm{o}$ with applied bias for $E_\mathrm{F}=0.2$ eV (blue curves) and $E_\mathrm{F}=0.4$ eV (red curves), where solid curves (left $y$-axis) correspond to real parts and dashed curves correspond to imaginary parts (right $y$-axis).}
\end{figure}

{To further illustrate the epsilon-near-zero shifting and the resonant nature of the in-plane dielectric response of this metamaterial, i.e. $\epsilon_\mathrm{o}$, in Fig. ~\ref{fig:epsilonFig} we show the relative change in dielectric permittivity, i.e. $\Delta\epsilon=100\times(\epsilon_\mathrm{o,V=0}-\epsilon_\mathrm{o,V})/\epsilon_\mathrm{o,V=0}$, for two different applied bias corresponding to $E_\mathrm{F=0.2}$ eV and to $E_\mathrm{F=0.4}$ eV, with blue and red color, respectively. These calculations were performed using the experimentally derived values for the optical properties and thicknesses of the constitutive components of the metamaterial, as described above. Near the surface phonon resonance of SiO$_\mathrm{2}$ at $20$ $\mu$m, significant tuning of the real part of $\epsilon_\mathrm{o}$ is observed, coming from the epsilon-near-zero tuning, which shifts by approximately $1$ micron. Bearing in mind that the out-of-plane response of this metamaterial ($\epsilon_\mathrm{e}$ in Eq. \ref{eq:1}) is not tunable due to the two-dimensional nature of graphene, as explained above, the change in sign of $\epsilon_\mathrm{o}$ on the left axis in Fig. ~\ref{fig:epsilonFig} corresponds to a topological transition of the isofrequency surface of this metamaterial.}

\par{In summary, we have experimentally demonstrated a graphene/polaritonic dielectric metamaterial with tunable epsilon-near-zero permittivity response. By tuning the Fermi level of graphene by $0.5$ eV, we observe a shift of $1$ $\mu$m in the near-zero response. Although previous theoretical proposals have focused on non-dispersive dielectric materials between graphene monolayers, here we showed that utilizing the polar response of dielectrics at infrared frequencies benefits tunability, and additionally provides means of tuning constitutive material properties of polar dielectrics and semiconductors, by incorporating graphene. Ellipsometry was used to determine the optical properties (dielectric response and thickness) of the constituent materials, and, based on effective parameter retrievals that homogenize the metamaterial, we experimentally characterized the full metamaterial stack. FTIR transmission measurements agree with our ellipsometric results, where transmission reduction is directly attributed to electrostatically induced charges in graphene and to epsilon-near-zero tuning.}

\begin{acknowledgments}

This work was supported by U.S. Department of Energy (DOE) Office of Science Grant No. DE-FG02-07ER46405 (G.T.P. and H.A.A.).  JB acknowledges support from a National Science Foundation Graduate Research Fellowship under Grant No. 1144469. G.T.P. acknowledges support from the TomKat Postdoctoral Fellowship in Sustainable Energy at Stanford University.

\end{acknowledgments}

%\nocite{*}
%\bibliography{aipsamp}% Produces the bibliography via BibTeX.

\begin{thebibliography}{64}%
\makeatletter
\providecommand \@ifxundefined [1]{%
 \@ifx{#1\undefined}
}%
\providecommand \@ifnum [1]{%
 \ifnum #1\expandafter \@firstoftwo
 \else \expandafter \@secondoftwo
 \fi
}%
\providecommand \@ifx [1]{%
 \ifx #1\expandafter \@firstoftwo
 \else \expandafter \@secondoftwo
 \fi
}%
\providecommand \natexlab [1]{#1}%
\providecommand \enquote  [1]{``#1''}%
\providecommand \bibnamefont  [1]{#1}%
\providecommand \bibfnamefont [1]{#1}%
\providecommand \citenamefont [1]{#1}%
\providecommand \href@noop [0]{\@secondoftwo}%
\providecommand \href [0]{\begingroup \@sanitize@url \@href}%
\providecommand \@href[1]{\@@startlink{#1}\@@href}%
\providecommand \@@href[1]{\endgroup#1\@@endlink}%
\providecommand \@sanitize@url [0]{\catcode `\\12\catcode `\$12\catcode
  `\&12\catcode `\#12\catcode `\^12\catcode `\_12\catcode `\%12\relax}%
\providecommand \@@startlink[1]{}%
\providecommand \@@endlink[0]{}%
\providecommand \url  [0]{\begingroup\@sanitize@url \@url }%
\providecommand \@url [1]{\endgroup\@href {#1}{\urlprefix }}%
\providecommand \urlprefix  [0]{URL }%
\providecommand \Eprint [0]{\href }%
\providecommand \doibase [0]{http://dx.doi.org/}%
\providecommand \selectlanguage [0]{\@gobble}%
\providecommand \bibinfo  [0]{\@secondoftwo}%
\providecommand \bibfield  [0]{\@secondoftwo}%
\providecommand \translation [1]{[#1]}%
\providecommand \BibitemOpen [0]{}%
\providecommand \bibitemStop [0]{}%
\providecommand \bibitemNoStop [0]{.\EOS\space}%
\providecommand \EOS [0]{\spacefactor3000\relax}%
\providecommand \BibitemShut  [1]{\csname bibitem#1\endcsname}%
\let\auto@bib@innerbib\@empty
%</preamble>
\bibitem [{\citenamefont {Lu}\ \emph {et~al.}(2014)\citenamefont {Lu},
  \citenamefont {Kan}, \citenamefont {Fullerton},\ and\ \citenamefont
  {Liu}}]{lu2014enhancing}%
  \BibitemOpen
  \bibfield  {author} {\bibinfo {author} {\bibfnamefont {D.}~\bibnamefont
  {Lu}}, \bibinfo {author} {\bibfnamefont {J.~J.}\ \bibnamefont {Kan}},
  \bibinfo {author} {\bibfnamefont {E.~E.}\ \bibnamefont {Fullerton}}, \ and\
  \bibinfo {author} {\bibfnamefont {Z.}~\bibnamefont {Liu}},\ }\bibfield
  {title} {\enquote {\bibinfo {title} {Enhancing spontaneous emission rates of
  molecules using nanopatterned multilayer hyperbolic metamaterials},}\
  }\href@noop {} {\bibfield  {journal} {\bibinfo  {journal} {Nature
  nanotechnology}\ }\textbf {\bibinfo {volume} {9}},\ \bibinfo {pages} {48}
  (\bibinfo {year} {2014})}\BibitemShut {NoStop}%
\bibitem [{\citenamefont {Zhou}\ \emph {et~al.}(2014)\citenamefont {Zhou},
  \citenamefont {Kaplan}, \citenamefont {Chen},\ and\ \citenamefont
  {Guo}}]{zhou2014experiment}%
  \BibitemOpen
  \bibfield  {author} {\bibinfo {author} {\bibfnamefont {J.}~\bibnamefont
  {Zhou}}, \bibinfo {author} {\bibfnamefont {A.~F.}\ \bibnamefont {Kaplan}},
  \bibinfo {author} {\bibfnamefont {L.}~\bibnamefont {Chen}}, \ and\ \bibinfo
  {author} {\bibfnamefont {L.~J.}\ \bibnamefont {Guo}},\ }\bibfield  {title}
  {\enquote {\bibinfo {title} {Experiment and theory of the broadband
  absorption by a tapered hyperbolic metamaterial array},}\ }\href@noop {}
  {\bibfield  {journal} {\bibinfo  {journal} {ACS photonics}\ }\textbf
  {\bibinfo {volume} {1}},\ \bibinfo {pages} {618--624} (\bibinfo {year}
  {2014})}\BibitemShut {NoStop}%
\bibitem [{\citenamefont {Jacob}\ \emph {et~al.}(2010)\citenamefont {Jacob},
  \citenamefont {Kim}, \citenamefont {Naik}, \citenamefont {Boltasseva},
  \citenamefont {Narimanov},\ and\ \citenamefont
  {Shalaev}}]{jacob2010engineering}%
  \BibitemOpen
  \bibfield  {author} {\bibinfo {author} {\bibfnamefont {Z.}~\bibnamefont
  {Jacob}}, \bibinfo {author} {\bibfnamefont {J.-Y.}\ \bibnamefont {Kim}},
  \bibinfo {author} {\bibfnamefont {G.~V.}\ \bibnamefont {Naik}}, \bibinfo
  {author} {\bibfnamefont {A.}~\bibnamefont {Boltasseva}}, \bibinfo {author}
  {\bibfnamefont {E.~E.}\ \bibnamefont {Narimanov}}, \ and\ \bibinfo {author}
  {\bibfnamefont {V.~M.}\ \bibnamefont {Shalaev}},\ }\bibfield  {title}
  {\enquote {\bibinfo {title} {Engineering photonic density of states using
  metamaterials},}\ }\href@noop {} {\bibfield  {journal} {\bibinfo  {journal}
  {Applied physics B}\ }\textbf {\bibinfo {volume} {100}},\ \bibinfo {pages}
  {215--218} (\bibinfo {year} {2010})}\BibitemShut {NoStop}%
\bibitem [{\citenamefont {Schuller}\ \emph {et~al.}(2010)\citenamefont
  {Schuller}, \citenamefont {Barnard}, \citenamefont {Cai}, \citenamefont
  {Jun}, \citenamefont {White},\ and\ \citenamefont
  {Brongersma}}]{schuller2010plasmonics}%
  \BibitemOpen
  \bibfield  {author} {\bibinfo {author} {\bibfnamefont {J.~A.}\ \bibnamefont
  {Schuller}}, \bibinfo {author} {\bibfnamefont {E.~S.}\ \bibnamefont
  {Barnard}}, \bibinfo {author} {\bibfnamefont {W.}~\bibnamefont {Cai}},
  \bibinfo {author} {\bibfnamefont {Y.~C.}\ \bibnamefont {Jun}}, \bibinfo
  {author} {\bibfnamefont {J.~S.}\ \bibnamefont {White}}, \ and\ \bibinfo
  {author} {\bibfnamefont {M.~L.}\ \bibnamefont {Brongersma}},\ }\bibfield
  {title} {\enquote {\bibinfo {title} {Plasmonics for extreme light
  concentration and manipulation},}\ }\href@noop {} {\bibfield  {journal}
  {\bibinfo  {journal} {Nature materials}\ }\textbf {\bibinfo {volume} {9}},\
  \bibinfo {pages} {193} (\bibinfo {year} {2010})}\BibitemShut {NoStop}%
\bibitem [{\citenamefont {Smith}\ \emph {et~al.}(2000)\citenamefont {Smith},
  \citenamefont {Padilla}, \citenamefont {Vier}, \citenamefont {Nemat-Nasser},\
  and\ \citenamefont {Schultz}}]{smith2000composite}%
  \BibitemOpen
  \bibfield  {author} {\bibinfo {author} {\bibfnamefont {D.~R.}\ \bibnamefont
  {Smith}}, \bibinfo {author} {\bibfnamefont {W.~J.}\ \bibnamefont {Padilla}},
  \bibinfo {author} {\bibfnamefont {D.}~\bibnamefont {Vier}}, \bibinfo {author}
  {\bibfnamefont {S.~C.}\ \bibnamefont {Nemat-Nasser}}, \ and\ \bibinfo
  {author} {\bibfnamefont {S.}~\bibnamefont {Schultz}},\ }\bibfield  {title}
  {\enquote {\bibinfo {title} {Composite medium with simultaneously negative
  permeability and permittivity},}\ }\href@noop {} {\bibfield  {journal}
  {\bibinfo  {journal} {Physical review letters}\ }\textbf {\bibinfo {volume}
  {84}},\ \bibinfo {pages} {4184} (\bibinfo {year} {2000})}\BibitemShut
  {NoStop}%
\bibitem [{\citenamefont {Poddubny}\ \emph {et~al.}(2013)\citenamefont
  {Poddubny}, \citenamefont {Iorsh}, \citenamefont {Belov},\ and\ \citenamefont
  {Kivshar}}]{poddubny2013hyperbolic}%
  \BibitemOpen
  \bibfield  {author} {\bibinfo {author} {\bibfnamefont {A.}~\bibnamefont
  {Poddubny}}, \bibinfo {author} {\bibfnamefont {I.}~\bibnamefont {Iorsh}},
  \bibinfo {author} {\bibfnamefont {P.}~\bibnamefont {Belov}}, \ and\ \bibinfo
  {author} {\bibfnamefont {Y.}~\bibnamefont {Kivshar}},\ }\bibfield  {title}
  {\enquote {\bibinfo {title} {Hyperbolic metamaterials},}\ }\href@noop {}
  {\bibfield  {journal} {\bibinfo  {journal} {Nature Photonics}\ }\textbf
  {\bibinfo {volume} {7}},\ \bibinfo {pages} {948} (\bibinfo {year}
  {2013})}\BibitemShut {NoStop}%
\bibitem [{\citenamefont {Smith}\ and\ \citenamefont
  {Schurig}(2003)}]{smith2003electromagnetic}%
  \BibitemOpen
  \bibfield  {author} {\bibinfo {author} {\bibfnamefont {D.}~\bibnamefont
  {Smith}}\ and\ \bibinfo {author} {\bibfnamefont {D.}~\bibnamefont
  {Schurig}},\ }\bibfield  {title} {\enquote {\bibinfo {title} {Electromagnetic
  wave propagation in media with indefinite permittivity and permeability
  tensors},}\ }\href@noop {} {\bibfield  {journal} {\bibinfo  {journal}
  {Physical Review Letters}\ }\textbf {\bibinfo {volume} {90}},\ \bibinfo
  {pages} {077405} (\bibinfo {year} {2003})}\BibitemShut {NoStop}%
\bibitem [{\citenamefont {Papadakis}, \citenamefont {Yeh},\ and\ \citenamefont
  {Atwater}(2015)}]{papadakis2015retrieval}%
  \BibitemOpen
  \bibfield  {author} {\bibinfo {author} {\bibfnamefont {G.~T.}\ \bibnamefont
  {Papadakis}}, \bibinfo {author} {\bibfnamefont {P.}~\bibnamefont {Yeh}}, \
  and\ \bibinfo {author} {\bibfnamefont {H.~A.}\ \bibnamefont {Atwater}},\
  }\bibfield  {title} {\enquote {\bibinfo {title} {Retrieval of material
  parameters for uniaxial metamaterials},}\ }\href@noop {} {\bibfield
  {journal} {\bibinfo  {journal} {Physical Review B}\ }\textbf {\bibinfo
  {volume} {91}},\ \bibinfo {pages} {155406} (\bibinfo {year}
  {2015})}\BibitemShut {NoStop}%
\bibitem [{\citenamefont {Babicheva}(2017)}]{Babicheva_WG}%
  \BibitemOpen
  \bibfield  {author} {\bibinfo {author} {\bibfnamefont {V.~E.}\ \bibnamefont
  {Babicheva}},\ }\bibfield  {title} {\enquote {\bibinfo {title} {Long-range
  propagation of plasmon and phonon polaritons in hyperbolic-metamaterial
  waveguides},}\ }\href {\doibase 10.1088/2040-8986/aa94b1} {\bibfield
  {journal} {\bibinfo  {journal} {Journal of Optics}\ }\textbf {\bibinfo
  {volume} {19}},\ \bibinfo {pages} {124013} (\bibinfo {year}
  {2017})}\BibitemShut {NoStop}%
\bibitem [{\citenamefont {Cortes}, \citenamefont {Otten},\ and\ \citenamefont
  {Gray}(2019)}]{cortes2019ground}%
  \BibitemOpen
  \bibfield  {author} {\bibinfo {author} {\bibfnamefont {C.~L.}\ \bibnamefont
  {Cortes}}, \bibinfo {author} {\bibfnamefont {M.}~\bibnamefont {Otten}}, \
  and\ \bibinfo {author} {\bibfnamefont {S.~K.}\ \bibnamefont {Gray}},\
  }\bibfield  {title} {\enquote {\bibinfo {title} {Ground-state cooling enabled
  by critical coupling and dark entangled states},}\ }\href@noop {} {\bibfield
  {journal} {\bibinfo  {journal} {Physical Review B}\ }\textbf {\bibinfo
  {volume} {99}},\ \bibinfo {pages} {014107} (\bibinfo {year}
  {2019})}\BibitemShut {NoStop}%
\bibitem [{\citenamefont {Biehs}, \citenamefont {Tschikin},\ and\ \citenamefont
  {Ben-Abdallah}(2012)}]{Biehs_BB}%
  \BibitemOpen
  \bibfield  {author} {\bibinfo {author} {\bibfnamefont {S.-A.}\ \bibnamefont
  {Biehs}}, \bibinfo {author} {\bibfnamefont {M.}~\bibnamefont {Tschikin}}, \
  and\ \bibinfo {author} {\bibfnamefont {P.}~\bibnamefont {Ben-Abdallah}},\
  }\bibfield  {title} {\enquote {\bibinfo {title} {Hyperbolic metamaterials as
  an analog of a blackbody in the near field},}\ }\href {\doibase
  10.1103/PhysRevLett.109.104301} {\bibfield  {journal} {\bibinfo  {journal}
  {Phys. Rev. Lett.}\ }\textbf {\bibinfo {volume} {109}},\ \bibinfo {pages}
  {104301} (\bibinfo {year} {2012})}\BibitemShut {NoStop}%
\bibitem [{\citenamefont {Fang}, \citenamefont {Koschny},\ and\ \citenamefont
  {Soukoulis}(2010)}]{fang2010lasing}%
  \BibitemOpen
  \bibfield  {author} {\bibinfo {author} {\bibfnamefont {A.}~\bibnamefont
  {Fang}}, \bibinfo {author} {\bibfnamefont {T.}~\bibnamefont {Koschny}}, \
  and\ \bibinfo {author} {\bibfnamefont {C.~M.}\ \bibnamefont {Soukoulis}},\
  }\bibfield  {title} {\enquote {\bibinfo {title} {Lasing in metamaterial
  nanostructures},}\ }\href@noop {} {\bibfield  {journal} {\bibinfo  {journal}
  {Journal of optics}\ }\textbf {\bibinfo {volume} {12}},\ \bibinfo {pages}
  {024013} (\bibinfo {year} {2010})}\BibitemShut {NoStop}%
\bibitem [{\citenamefont {Liu}\ \emph {et~al.}(2007)\citenamefont {Liu},
  \citenamefont {Lee}, \citenamefont {Xiong}, \citenamefont {Sun},\ and\
  \citenamefont {Zhang}}]{Zhang_Science}%
  \BibitemOpen
  \bibfield  {author} {\bibinfo {author} {\bibfnamefont {Z.}~\bibnamefont
  {Liu}}, \bibinfo {author} {\bibfnamefont {H.}~\bibnamefont {Lee}}, \bibinfo
  {author} {\bibfnamefont {Y.}~\bibnamefont {Xiong}}, \bibinfo {author}
  {\bibfnamefont {C.}~\bibnamefont {Sun}}, \ and\ \bibinfo {author}
  {\bibfnamefont {X.}~\bibnamefont {Zhang}},\ }\bibfield  {title} {\enquote
  {\bibinfo {title} {Far-field optical hyperlens magnifying
  sub-diffraction-limited objects},}\ }\href {\doibase 10.1126/science.1137368}
  {\bibfield  {journal} {\bibinfo  {journal} {Science}\ }\textbf {\bibinfo
  {volume} {315}},\ \bibinfo {pages} {1686--1686} (\bibinfo {year}
  {2007})}\BibitemShut {NoStop}%
\bibitem [{\citenamefont {Weile}(2007)}]{weile2007electromagnetic}%
  \BibitemOpen
  \bibfield  {author} {\bibinfo {author} {\bibfnamefont {D.~S.}\ \bibnamefont
  {Weile}},\ }\bibfield  {title} {\enquote {\bibinfo {title} {Electromagnetic
  metamaterials: Physics and engineering explorations (engheta, n. and
  ziolkowski, rw; 2006)[book review]},}\ }\href@noop {} {\bibfield  {journal}
  {\bibinfo  {journal} {IEEE Antennas and Propagation Magazine}\ }\textbf
  {\bibinfo {volume} {49}},\ \bibinfo {pages} {137--139} (\bibinfo {year}
  {2007})}\BibitemShut {NoStop}%
\bibitem [{\citenamefont {Maas}\ \emph {et~al.}(2013)\citenamefont {Maas},
  \citenamefont {Parsons}, \citenamefont {Engheta},\ and\ \citenamefont
  {Polman}}]{maas2013experimental}%
  \BibitemOpen
  \bibfield  {author} {\bibinfo {author} {\bibfnamefont {R.}~\bibnamefont
  {Maas}}, \bibinfo {author} {\bibfnamefont {J.}~\bibnamefont {Parsons}},
  \bibinfo {author} {\bibfnamefont {N.}~\bibnamefont {Engheta}}, \ and\
  \bibinfo {author} {\bibfnamefont {A.}~\bibnamefont {Polman}},\ }\bibfield
  {title} {\enquote {\bibinfo {title} {Experimental realization of an
  epsilon-near-zero metamaterial at visible wavelengths},}\ }\href@noop {}
  {\bibfield  {journal} {\bibinfo  {journal} {Nature Photonics}\ }\textbf
  {\bibinfo {volume} {7}},\ \bibinfo {pages} {907} (\bibinfo {year}
  {2013})}\BibitemShut {NoStop}%
\bibitem [{\citenamefont {Mahmoud}\ and\ \citenamefont
  {Engheta}(2014)}]{mahmoud2014wave}%
  \BibitemOpen
  \bibfield  {author} {\bibinfo {author} {\bibfnamefont {A.~M.}\ \bibnamefont
  {Mahmoud}}\ and\ \bibinfo {author} {\bibfnamefont {N.}~\bibnamefont
  {Engheta}},\ }\bibfield  {title} {\enquote {\bibinfo {title} {Wave--matter
  interactions in epsilon-and-mu-near-zero structures},}\ }\href@noop {}
  {\bibfield  {journal} {\bibinfo  {journal} {Nature Communications}\ }\textbf
  {\bibinfo {volume} {5}},\ \bibinfo {pages} {5638} (\bibinfo {year}
  {2014})}\BibitemShut {NoStop}%
\bibitem [{\citenamefont {Engheta}(2013)}]{engheta2013pursuing}%
  \BibitemOpen
  \bibfield  {author} {\bibinfo {author} {\bibfnamefont {N.}~\bibnamefont
  {Engheta}},\ }\bibfield  {title} {\enquote {\bibinfo {title} {Pursuing
  near-zero response},}\ }\href@noop {} {\bibfield  {journal} {\bibinfo
  {journal} {Science}\ }\textbf {\bibinfo {volume} {340}},\ \bibinfo {pages}
  {286--287} (\bibinfo {year} {2013})}\BibitemShut {NoStop}%
\bibitem [{\citenamefont {Papadakis}\ and\ \citenamefont
  {Atwater}(2015)}]{papadakis_tunable}%
  \BibitemOpen
  \bibfield  {author} {\bibinfo {author} {\bibfnamefont {G.~T.}\ \bibnamefont
  {Papadakis}}\ and\ \bibinfo {author} {\bibfnamefont {H.~A.}\ \bibnamefont
  {Atwater}},\ }\bibfield  {title} {\enquote {\bibinfo {title} {Field-effect
  induced tunability in hyperbolic metamaterials},}\ }\href {\doibase
  10.1103/PhysRevB.92.184101} {\bibfield  {journal} {\bibinfo  {journal} {Phys.
  Rev. B}\ }\textbf {\bibinfo {volume} {92}},\ \bibinfo {pages} {184101}
  (\bibinfo {year} {2015})}\BibitemShut {NoStop}%
\bibitem [{\citenamefont {Roberts}\ \emph {et~al.}(2019)\citenamefont
  {Roberts}, \citenamefont {Yu}, \citenamefont {Ho}, \citenamefont {Schoeche},
  \citenamefont {Falk},\ and\ \citenamefont {Fan}}]{FanJon_HMM}%
  \BibitemOpen
  \bibfield  {author} {\bibinfo {author} {\bibfnamefont {J.~A.}\ \bibnamefont
  {Roberts}}, \bibinfo {author} {\bibfnamefont {S.-J.}\ \bibnamefont {Yu}},
  \bibinfo {author} {\bibfnamefont {P.-H.}\ \bibnamefont {Ho}}, \bibinfo
  {author} {\bibfnamefont {S.}~\bibnamefont {Schoeche}}, \bibinfo {author}
  {\bibfnamefont {A.~L.}\ \bibnamefont {Falk}}, \ and\ \bibinfo {author}
  {\bibfnamefont {J.~A.}\ \bibnamefont {Fan}},\ }\bibfield  {title} {\enquote
  {\bibinfo {title} {Tunable hyperbolic metamaterials based on self-assembled
  carbon nanotubes},}\ }\href {\doibase 10.1021/acs.nanolett.9b00552}
  {\bibfield  {journal} {\bibinfo  {journal} {Nano Letters}\ }\textbf {\bibinfo
  {volume} {19}},\ \bibinfo {pages} {3131} (\bibinfo {year}
  {2019})}\BibitemShut {NoStop}%
\bibitem [{\citenamefont {Lu}, \citenamefont {Simpson},\ and\ \citenamefont
  {Valiyaveedu}(2018)}]{Tunable_HMM_review}%
  \BibitemOpen
  \bibfield  {author} {\bibinfo {author} {\bibfnamefont {L.}~\bibnamefont
  {Lu}}, \bibinfo {author} {\bibfnamefont {R.~E.}\ \bibnamefont {Simpson}}, \
  and\ \bibinfo {author} {\bibfnamefont {S.~K.}\ \bibnamefont {Valiyaveedu}},\
  }\bibfield  {title} {\enquote {\bibinfo {title} {Active hyperbolic
  metamaterials: progress, materials and design},}\ }\href@noop {} {\bibfield
  {journal} {\bibinfo  {journal} {Journal of Optics}\ }\textbf {\bibinfo
  {volume} {10}},\ \bibinfo {pages} {103001} (\bibinfo {year}
  {2018})}\BibitemShut {NoStop}%
\bibitem [{\citenamefont {Novoselov}\ \emph {et~al.}(2005)\citenamefont
  {Novoselov}, \citenamefont {Geim}, \citenamefont {Morozov}, \citenamefont
  {Jiang}, \citenamefont {Katsnelson}, \citenamefont {Grigorieva},
  \citenamefont {Dubonos}, \citenamefont {Firsov},\ and\ \citenamefont
  {AA}}]{novoselov2005two}%
  \BibitemOpen
  \bibfield  {author} {\bibinfo {author} {\bibfnamefont {K.~S.}\ \bibnamefont
  {Novoselov}}, \bibinfo {author} {\bibfnamefont {A.~K.}\ \bibnamefont {Geim}},
  \bibinfo {author} {\bibfnamefont {S.}~\bibnamefont {Morozov}}, \bibinfo
  {author} {\bibfnamefont {D.}~\bibnamefont {Jiang}}, \bibinfo {author}
  {\bibfnamefont {M.}~\bibnamefont {Katsnelson}}, \bibinfo {author}
  {\bibfnamefont {I.}~\bibnamefont {Grigorieva}}, \bibinfo {author}
  {\bibfnamefont {S.}~\bibnamefont {Dubonos}}, \bibinfo {author} {\bibnamefont
  {Firsov}}, \ and\ \bibinfo {author} {\bibnamefont {AA}},\ }\bibfield  {title}
  {\enquote {\bibinfo {title} {Two-dimensional gas of massless dirac fermions
  in graphene},}\ }\href@noop {} {\bibfield  {journal} {\bibinfo  {journal}
  {nature}\ }\textbf {\bibinfo {volume} {438}},\ \bibinfo {pages} {197}
  (\bibinfo {year} {2005})}\BibitemShut {NoStop}%
\bibitem [{\citenamefont {Andersen}(2010)}]{andersen2010graphene}%
  \BibitemOpen
  \bibfield  {author} {\bibinfo {author} {\bibfnamefont {D.~R.}\ \bibnamefont
  {Andersen}},\ }\bibfield  {title} {\enquote {\bibinfo {title} {Graphene-based
  long-wave infrared tm surface plasmon modulator},}\ }\href@noop {} {\bibfield
   {journal} {\bibinfo  {journal} {JOSA B}\ }\textbf {\bibinfo {volume} {27}},\
  \bibinfo {pages} {818--823} (\bibinfo {year} {2010})}\BibitemShut {NoStop}%
\bibitem [{\citenamefont {Ryzhii}, \citenamefont {Ryzhii},\ and\ \citenamefont
  {Otsuji}(2007)}]{ryzhii2007negative}%
  \BibitemOpen
  \bibfield  {author} {\bibinfo {author} {\bibfnamefont {V.}~\bibnamefont
  {Ryzhii}}, \bibinfo {author} {\bibfnamefont {M.}~\bibnamefont {Ryzhii}}, \
  and\ \bibinfo {author} {\bibfnamefont {T.}~\bibnamefont {Otsuji}},\
  }\bibfield  {title} {\enquote {\bibinfo {title} {Negative dynamic
  conductivity of graphene with optical pumping},}\ }\href@noop {} {\bibfield
  {journal} {\bibinfo  {journal} {Journal of Applied Physics}\ }\textbf
  {\bibinfo {volume} {101}},\ \bibinfo {pages} {083114} (\bibinfo {year}
  {2007})}\BibitemShut {NoStop}%
\bibitem [{\citenamefont {Vakil}\ and\ \citenamefont
  {Engheta}(2011)}]{vakil2011transformation}%
  \BibitemOpen
  \bibfield  {author} {\bibinfo {author} {\bibfnamefont {A.}~\bibnamefont
  {Vakil}}\ and\ \bibinfo {author} {\bibfnamefont {N.}~\bibnamefont
  {Engheta}},\ }\bibfield  {title} {\enquote {\bibinfo {title} {Transformation
  optics using graphene},}\ }\href@noop {} {\bibfield  {journal} {\bibinfo
  {journal} {Science}\ }\textbf {\bibinfo {volume} {332}},\ \bibinfo {pages}
  {1291--1294} (\bibinfo {year} {2011})}\BibitemShut {NoStop}%
\bibitem [{\citenamefont {Polini}\ \emph {et~al.}(2008)\citenamefont {Polini},
  \citenamefont {Asgari}, \citenamefont {Borghi}, \citenamefont {Barlas},
  \citenamefont {Pereg-Barnea},\ and\ \citenamefont
  {MacDonald}}]{polini2008plasmons}%
  \BibitemOpen
  \bibfield  {author} {\bibinfo {author} {\bibfnamefont {M.}~\bibnamefont
  {Polini}}, \bibinfo {author} {\bibfnamefont {R.}~\bibnamefont {Asgari}},
  \bibinfo {author} {\bibfnamefont {G.}~\bibnamefont {Borghi}}, \bibinfo
  {author} {\bibfnamefont {Y.}~\bibnamefont {Barlas}}, \bibinfo {author}
  {\bibfnamefont {T.}~\bibnamefont {Pereg-Barnea}}, \ and\ \bibinfo {author}
  {\bibfnamefont {A.}~\bibnamefont {MacDonald}},\ }\bibfield  {title} {\enquote
  {\bibinfo {title} {Plasmons and the spectral function of graphene},}\
  }\href@noop {} {\bibfield  {journal} {\bibinfo  {journal} {Physical Review
  B}\ }\textbf {\bibinfo {volume} {77}},\ \bibinfo {pages} {081411} (\bibinfo
  {year} {2008})}\BibitemShut {NoStop}%
\bibitem [{\citenamefont {Hwang}\ and\ \citenamefont
  {Sarma}(2007)}]{hwang2007dielectric}%
  \BibitemOpen
  \bibfield  {author} {\bibinfo {author} {\bibfnamefont {E.}~\bibnamefont
  {Hwang}}\ and\ \bibinfo {author} {\bibfnamefont {S.~D.}\ \bibnamefont
  {Sarma}},\ }\bibfield  {title} {\enquote {\bibinfo {title} {Dielectric
  function, screening, and plasmons in two-dimensional graphene},}\ }\href@noop
  {} {\bibfield  {journal} {\bibinfo  {journal} {Physical Review B}\ }\textbf
  {\bibinfo {volume} {75}},\ \bibinfo {pages} {205418} (\bibinfo {year}
  {2007})}\BibitemShut {NoStop}%
\bibitem [{\citenamefont {Brar}\ \emph {et~al.}(2013)\citenamefont {Brar},
  \citenamefont {Jang}, \citenamefont {Sherrott}, \citenamefont {Lopez},\ and\
  \citenamefont {Atwater}}]{brar2013highly}%
  \BibitemOpen
  \bibfield  {author} {\bibinfo {author} {\bibfnamefont {V.~W.}\ \bibnamefont
  {Brar}}, \bibinfo {author} {\bibfnamefont {M.~S.}\ \bibnamefont {Jang}},
  \bibinfo {author} {\bibfnamefont {M.}~\bibnamefont {Sherrott}}, \bibinfo
  {author} {\bibfnamefont {J.~J.}\ \bibnamefont {Lopez}}, \ and\ \bibinfo
  {author} {\bibfnamefont {H.~A.}\ \bibnamefont {Atwater}},\ }\bibfield
  {title} {\enquote {\bibinfo {title} {Highly confined tunable mid-infrared
  plasmonics in graphene nanoresonators},}\ }\href@noop {} {\bibfield
  {journal} {\bibinfo  {journal} {Nano letters}\ }\textbf {\bibinfo {volume}
  {13}},\ \bibinfo {pages} {2541--2547} (\bibinfo {year} {2013})}\BibitemShut
  {NoStop}%
\bibitem [{\citenamefont {Linder}\ and\ \citenamefont
  {Halterman}(2016)}]{linder2016graphene}%
  \BibitemOpen
  \bibfield  {author} {\bibinfo {author} {\bibfnamefont {J.}~\bibnamefont
  {Linder}}\ and\ \bibinfo {author} {\bibfnamefont {K.}~\bibnamefont
  {Halterman}},\ }\bibfield  {title} {\enquote {\bibinfo {title}
  {Graphene-based extremely wide-angle tunable metamaterial absorber},}\
  }\href@noop {} {\bibfield  {journal} {\bibinfo  {journal} {Scientific
  reports}\ }\textbf {\bibinfo {volume} {6}},\ \bibinfo {pages} {31225}
  (\bibinfo {year} {2016})}\BibitemShut {NoStop}%
\bibitem [{\citenamefont {Andryieuski}\ and\ \citenamefont
  {Lavrinenko}(2013)}]{andryieuski2013graphene}%
  \BibitemOpen
  \bibfield  {author} {\bibinfo {author} {\bibfnamefont {A.}~\bibnamefont
  {Andryieuski}}\ and\ \bibinfo {author} {\bibfnamefont {A.~V.}\ \bibnamefont
  {Lavrinenko}},\ }\bibfield  {title} {\enquote {\bibinfo {title} {Graphene
  metamaterials based tunable terahertz absorber: effective surface
  conductivity approach},}\ }\href@noop {} {\bibfield  {journal} {\bibinfo
  {journal} {Optics express}\ }\textbf {\bibinfo {volume} {21}},\ \bibinfo
  {pages} {9144--9155} (\bibinfo {year} {2013})}\BibitemShut {NoStop}%
\bibitem [{\citenamefont {Xiong}\ \emph {et~al.}(2018)\citenamefont {Xiong},
  \citenamefont {Wu}, \citenamefont {Dong}, \citenamefont {Tang}, \citenamefont
  {Jiang},\ and\ \citenamefont {Zeng}}]{xiong2018ultra}%
  \BibitemOpen
  \bibfield  {author} {\bibinfo {author} {\bibfnamefont {H.}~\bibnamefont
  {Xiong}}, \bibinfo {author} {\bibfnamefont {Y.-B.}\ \bibnamefont {Wu}},
  \bibinfo {author} {\bibfnamefont {J.}~\bibnamefont {Dong}}, \bibinfo {author}
  {\bibfnamefont {M.-C.}\ \bibnamefont {Tang}}, \bibinfo {author}
  {\bibfnamefont {Y.-N.}\ \bibnamefont {Jiang}}, \ and\ \bibinfo {author}
  {\bibfnamefont {X.-P.}\ \bibnamefont {Zeng}},\ }\bibfield  {title} {\enquote
  {\bibinfo {title} {Ultra-thin and broadband tunable metamaterial graphene
  absorber},}\ }\href@noop {} {\bibfield  {journal} {\bibinfo  {journal}
  {Optics express}\ }\textbf {\bibinfo {volume} {26}},\ \bibinfo {pages}
  {1681--1688} (\bibinfo {year} {2018})}\BibitemShut {NoStop}%
\bibitem [{\citenamefont {Zainud-Deen}, \citenamefont {Mabrouk},\ and\
  \citenamefont {Malhat}(2017)}]{zainud2017frequency}%
  \BibitemOpen
  \bibfield  {author} {\bibinfo {author} {\bibfnamefont {S.}~\bibnamefont
  {Zainud-Deen}}, \bibinfo {author} {\bibfnamefont {A.}~\bibnamefont
  {Mabrouk}}, \ and\ \bibinfo {author} {\bibfnamefont {H.}~\bibnamefont
  {Malhat}},\ }\bibfield  {title} {\enquote {\bibinfo {title} {Frequency
  tunable graphene metamaterial reflectarray},}\ }in\ \href@noop {} {\emph
  {\bibinfo {booktitle} {2017 XXXIInd General Assembly and Scientific Symposium
  of the International Union of Radio Science (URSI GASS)}}}\ (\bibinfo
  {organization} {IEEE},\ \bibinfo {year} {2017})\ pp.\ \bibinfo {pages}
  {1--4}\BibitemShut {NoStop}%
\bibitem [{\citenamefont {Al~Sayem}\ \emph {et~al.}(2014)\citenamefont
  {Al~Sayem}, \citenamefont {Shahriar}, \citenamefont {Mahdy},\ and\
  \citenamefont {Rahman}}]{al2014control}%
  \BibitemOpen
  \bibfield  {author} {\bibinfo {author} {\bibfnamefont {A.}~\bibnamefont
  {Al~Sayem}}, \bibinfo {author} {\bibfnamefont {A.}~\bibnamefont {Shahriar}},
  \bibinfo {author} {\bibfnamefont {M.}~\bibnamefont {Mahdy}}, \ and\ \bibinfo
  {author} {\bibfnamefont {M.~S.}\ \bibnamefont {Rahman}},\ }\bibfield  {title}
  {\enquote {\bibinfo {title} {Control of reflection through epsilon near zero
  graphene based anisotropic metamaterial},}\ }in\ \href@noop {} {\emph
  {\bibinfo {booktitle} {8th International Conference on Electrical and
  Computer Engineering}}}\ (\bibinfo {organization} {IEEE},\ \bibinfo {year}
  {2014})\ pp.\ \bibinfo {pages} {812--815}\BibitemShut {NoStop}%
\bibitem [{\citenamefont {Iorsh}\ \emph {et~al.}(2013)\citenamefont {Iorsh},
  \citenamefont {Mukhin}, \citenamefont {Shadrivov}, \citenamefont {Belov},\
  and\ \citenamefont {Kivshar}}]{iorsh2013hyperbolic}%
  \BibitemOpen
  \bibfield  {author} {\bibinfo {author} {\bibfnamefont {I.~V.}\ \bibnamefont
  {Iorsh}}, \bibinfo {author} {\bibfnamefont {I.~S.}\ \bibnamefont {Mukhin}},
  \bibinfo {author} {\bibfnamefont {I.~V.}\ \bibnamefont {Shadrivov}}, \bibinfo
  {author} {\bibfnamefont {P.~A.}\ \bibnamefont {Belov}}, \ and\ \bibinfo
  {author} {\bibfnamefont {Y.~S.}\ \bibnamefont {Kivshar}},\ }\bibfield
  {title} {\enquote {\bibinfo {title} {Hyperbolic metamaterials based on
  multilayer graphene structures},}\ }\href@noop {} {\bibfield  {journal}
  {\bibinfo  {journal} {Physical Review B}\ }\textbf {\bibinfo {volume} {87}},\
  \bibinfo {pages} {075416} (\bibinfo {year} {2013})}\BibitemShut {NoStop}%
\bibitem [{\citenamefont {Othman}, \citenamefont {Guclu},\ and\ \citenamefont
  {Capolino}(2013)}]{othman2013graphene}%
  \BibitemOpen
  \bibfield  {author} {\bibinfo {author} {\bibfnamefont {M.~A.}\ \bibnamefont
  {Othman}}, \bibinfo {author} {\bibfnamefont {C.}~\bibnamefont {Guclu}}, \
  and\ \bibinfo {author} {\bibfnamefont {F.}~\bibnamefont {Capolino}},\
  }\bibfield  {title} {\enquote {\bibinfo {title} {Graphene--dielectric
  composite metamaterials: evolution from elliptic to hyperbolic wavevector
  dispersion and the transverse epsilon-near-zero condition},}\ }\href@noop {}
  {\bibfield  {journal} {\bibinfo  {journal} {Journal of Nanophotonics}\
  }\textbf {\bibinfo {volume} {7}},\ \bibinfo {pages} {073089} (\bibinfo {year}
  {2013})}\BibitemShut {NoStop}%
\bibitem [{\citenamefont {Janaszek}, \citenamefont {Tyszka-Zawadzka},\ and\
  \citenamefont {Szczepa\'{n}ski}(2016)}]{Janaszek2016}%
  \BibitemOpen
  \bibfield  {author} {\bibinfo {author} {\bibfnamefont {B.}~\bibnamefont
  {Janaszek}}, \bibinfo {author} {\bibfnamefont {A.}~\bibnamefont
  {Tyszka-Zawadzka}}, \ and\ \bibinfo {author} {\bibfnamefont {P.}~\bibnamefont
  {Szczepa\'{n}ski}},\ }\bibfield  {title} {\enquote {\bibinfo {title} {Tunable
  graphene-based hyperbolic metamaterial operating in sclu telecom bands},}\
  }\href@noop {} {\bibfield  {journal} {\bibinfo  {journal} {Optics express}\
  }\textbf {\bibinfo {volume} {24}},\ \bibinfo {pages} {24129--24136} (\bibinfo
  {year} {2016})}\BibitemShut {NoStop}%
\bibitem [{\citenamefont {Chang}\ \emph {et~al.}(2016)\citenamefont {Chang},
  \citenamefont {Liu}, \citenamefont {Liu}, \citenamefont {Zhang},
  \citenamefont {Marder}, \citenamefont {Narimanov}, \citenamefont {Zhong},\
  and\ \citenamefont {Norris}}]{chang2016realization}%
  \BibitemOpen
  \bibfield  {author} {\bibinfo {author} {\bibfnamefont {Y.-C.}\ \bibnamefont
  {Chang}}, \bibinfo {author} {\bibfnamefont {C.-H.}\ \bibnamefont {Liu}},
  \bibinfo {author} {\bibfnamefont {C.-H.}\ \bibnamefont {Liu}}, \bibinfo
  {author} {\bibfnamefont {S.}~\bibnamefont {Zhang}}, \bibinfo {author}
  {\bibfnamefont {S.~R.}\ \bibnamefont {Marder}}, \bibinfo {author}
  {\bibfnamefont {E.~E.}\ \bibnamefont {Narimanov}}, \bibinfo {author}
  {\bibfnamefont {Z.}~\bibnamefont {Zhong}}, \ and\ \bibinfo {author}
  {\bibfnamefont {T.~B.}\ \bibnamefont {Norris}},\ }\bibfield  {title}
  {\enquote {\bibinfo {title} {Realization of mid-infrared graphene hyperbolic
  metamaterials},}\ }\href@noop {} {\bibfield  {journal} {\bibinfo  {journal}
  {Nature communications}\ }\textbf {\bibinfo {volume} {7}},\ \bibinfo {pages}
  {10568} (\bibinfo {year} {2016})}\BibitemShut {NoStop}%
\bibitem [{\citenamefont {Dai}\ \emph {et~al.}(2015{\natexlab{a}})\citenamefont
  {Dai}, \citenamefont {Ma}, \citenamefont {Liu}, \citenamefont {Andersen},
  \citenamefont {Fei}, \citenamefont {Goldflam}, \citenamefont {Wagner},
  \citenamefont {Watanabe}, \citenamefont {Taniguchi}, \citenamefont
  {Thiemens}, \citenamefont {Keilmann}, \citenamefont {Janssen}, \citenamefont
  {Zhu}, \citenamefont {Jarillo-Herrero}, \citenamefont {Fogler},\ and\
  \citenamefont {Basov}}]{Basov_GHMM}%
  \BibitemOpen
  \bibfield  {author} {\bibinfo {author} {\bibfnamefont {S.}~\bibnamefont
  {Dai}}, \bibinfo {author} {\bibfnamefont {Q.}~\bibnamefont {Ma}}, \bibinfo
  {author} {\bibfnamefont {M.~K.}\ \bibnamefont {Liu}}, \bibinfo {author}
  {\bibfnamefont {T.}~\bibnamefont {Andersen}}, \bibinfo {author}
  {\bibfnamefont {Z.}~\bibnamefont {Fei}}, \bibinfo {author} {\bibfnamefont
  {M.~D.}\ \bibnamefont {Goldflam}}, \bibinfo {author} {\bibfnamefont
  {M.}~\bibnamefont {Wagner}}, \bibinfo {author} {\bibfnamefont
  {K.}~\bibnamefont {Watanabe}}, \bibinfo {author} {\bibfnamefont
  {T.}~\bibnamefont {Taniguchi}}, \bibinfo {author} {\bibfnamefont
  {M.}~\bibnamefont {Thiemens}}, \bibinfo {author} {\bibfnamefont
  {F.}~\bibnamefont {Keilmann}}, \bibinfo {author} {\bibfnamefont {G.~C.
  A.~M.}\ \bibnamefont {Janssen}}, \bibinfo {author} {\bibfnamefont {S.-E.}\
  \bibnamefont {Zhu}}, \bibinfo {author} {\bibfnamefont {P.}~\bibnamefont
  {Jarillo-Herrero}}, \bibinfo {author} {\bibfnamefont {M.~M.}\ \bibnamefont
  {Fogler}}, \ and\ \bibinfo {author} {\bibfnamefont {D.~N.}\ \bibnamefont
  {Basov}},\ }\bibfield  {title} {\enquote {\bibinfo {title} {Graphene on
  hexagonal boron nitride as a tunable hyperbolic metamaterial},}\ }\href@noop
  {} {\bibfield  {journal} {\bibinfo  {journal} {Nature nanotechnology}\
  }\textbf {\bibinfo {volume} {10}},\ \bibinfo {pages} {682} (\bibinfo {year}
  {2015}{\natexlab{a}})}\BibitemShut {NoStop}%
\bibitem [{\citenamefont {McNerny}\ \emph {et~al.}(2014)\citenamefont
  {McNerny}, \citenamefont {Viswanath}, \citenamefont {Copic}, \citenamefont
  {Laye}, \citenamefont {Prohoda}, \citenamefont {Brieland-Shoultz},
  \citenamefont {Polsen}, \citenamefont {Dee}, \citenamefont {Veerasamy},\ and\
  \citenamefont {Hart}}]{mcnerny2014direct}%
  \BibitemOpen
  \bibfield  {author} {\bibinfo {author} {\bibfnamefont {D.~Q.}\ \bibnamefont
  {McNerny}}, \bibinfo {author} {\bibfnamefont {B.}~\bibnamefont {Viswanath}},
  \bibinfo {author} {\bibfnamefont {D.}~\bibnamefont {Copic}}, \bibinfo
  {author} {\bibfnamefont {F.~R.}\ \bibnamefont {Laye}}, \bibinfo {author}
  {\bibfnamefont {C.}~\bibnamefont {Prohoda}}, \bibinfo {author} {\bibfnamefont
  {A.~C.}\ \bibnamefont {Brieland-Shoultz}}, \bibinfo {author} {\bibfnamefont
  {E.~S.}\ \bibnamefont {Polsen}}, \bibinfo {author} {\bibfnamefont {N.~T.}\
  \bibnamefont {Dee}}, \bibinfo {author} {\bibfnamefont {V.~S.}\ \bibnamefont
  {Veerasamy}}, \ and\ \bibinfo {author} {\bibfnamefont {A.~J.}\ \bibnamefont
  {Hart}},\ }\bibfield  {title} {\enquote {\bibinfo {title} {Direct fabrication
  of graphene on sio 2 enabled by thin film stress engineering},}\ }\href@noop
  {} {\bibfield  {journal} {\bibinfo  {journal} {Scientific reports}\ }\textbf
  {\bibinfo {volume} {4}},\ \bibinfo {pages} {5049} (\bibinfo {year}
  {2014})}\BibitemShut {NoStop}%
\bibitem [{\citenamefont {Park}\ \emph {et~al.}(2014)\citenamefont {Park},
  \citenamefont {Kihm}, \citenamefont {Kim}, \citenamefont {Lim}, \citenamefont
  {Cheon},\ and\ \citenamefont {Lee}}]{park2014wetting}%
  \BibitemOpen
  \bibfield  {author} {\bibinfo {author} {\bibfnamefont {J.~S.}\ \bibnamefont
  {Park}}, \bibinfo {author} {\bibfnamefont {K.~D.}\ \bibnamefont {Kihm}},
  \bibinfo {author} {\bibfnamefont {H.}~\bibnamefont {Kim}}, \bibinfo {author}
  {\bibfnamefont {G.}~\bibnamefont {Lim}}, \bibinfo {author} {\bibfnamefont
  {S.}~\bibnamefont {Cheon}}, \ and\ \bibinfo {author} {\bibfnamefont {J.~S.}\
  \bibnamefont {Lee}},\ }\bibfield  {title} {\enquote {\bibinfo {title}
  {Wetting and evaporative aggregation of nanofluid droplets on cvd-synthesized
  hydrophobic graphene surfaces},}\ }\href@noop {} {\bibfield  {journal}
  {\bibinfo  {journal} {Langmuir}\ }\textbf {\bibinfo {volume} {30}},\ \bibinfo
  {pages} {8268--8275} (\bibinfo {year} {2014})}\BibitemShut {NoStop}%
\bibitem [{\citenamefont {Lee}\ \emph {et~al.}(2010)\citenamefont {Lee},
  \citenamefont {Mordi}, \citenamefont {Kim}, \citenamefont {Chabal},
  \citenamefont {Vogel}, \citenamefont {Wallace}, \citenamefont {Cho},
  \citenamefont {Colombo},\ and\ \citenamefont {Kim}}]{lee2010characteristics}%
  \BibitemOpen
  \bibfield  {author} {\bibinfo {author} {\bibfnamefont {B.}~\bibnamefont
  {Lee}}, \bibinfo {author} {\bibfnamefont {G.}~\bibnamefont {Mordi}}, \bibinfo
  {author} {\bibfnamefont {M.}~\bibnamefont {Kim}}, \bibinfo {author}
  {\bibfnamefont {Y.}~\bibnamefont {Chabal}}, \bibinfo {author} {\bibfnamefont
  {E.}~\bibnamefont {Vogel}}, \bibinfo {author} {\bibfnamefont
  {R.}~\bibnamefont {Wallace}}, \bibinfo {author} {\bibfnamefont
  {K.}~\bibnamefont {Cho}}, \bibinfo {author} {\bibfnamefont {L.}~\bibnamefont
  {Colombo}}, \ and\ \bibinfo {author} {\bibfnamefont {J.}~\bibnamefont
  {Kim}},\ }\bibfield  {title} {\enquote {\bibinfo {title} {Characteristics of
  high-k al 2 o 3 dielectric using ozone-based atomic layer deposition for
  dual-gated graphene devices},}\ }\href@noop {} {\bibfield  {journal}
  {\bibinfo  {journal} {Applied Physics Letters}\ }\textbf {\bibinfo {volume}
  {97}},\ \bibinfo {pages} {043107} (\bibinfo {year} {2010})}\BibitemShut
  {NoStop}%
\bibitem [{\citenamefont {Kim}\ \emph {et~al.}(2009)\citenamefont {Kim},
  \citenamefont {Nah}, \citenamefont {Jo}, \citenamefont {Shahrjerdi},
  \citenamefont {Colombo}, \citenamefont {Yao}, \citenamefont {Tutuc},\ and\
  \citenamefont {Banerjee}}]{kim2009realization}%
  \BibitemOpen
  \bibfield  {author} {\bibinfo {author} {\bibfnamefont {S.}~\bibnamefont
  {Kim}}, \bibinfo {author} {\bibfnamefont {J.}~\bibnamefont {Nah}}, \bibinfo
  {author} {\bibfnamefont {I.}~\bibnamefont {Jo}}, \bibinfo {author}
  {\bibfnamefont {D.}~\bibnamefont {Shahrjerdi}}, \bibinfo {author}
  {\bibfnamefont {L.}~\bibnamefont {Colombo}}, \bibinfo {author} {\bibfnamefont
  {Z.}~\bibnamefont {Yao}}, \bibinfo {author} {\bibfnamefont {E.}~\bibnamefont
  {Tutuc}}, \ and\ \bibinfo {author} {\bibfnamefont {S.~K.}\ \bibnamefont
  {Banerjee}},\ }\bibfield  {title} {\enquote {\bibinfo {title} {Realization of
  a high mobility dual-gated graphene field-effect transistor with al 2 o 3
  dielectric},}\ }\href@noop {} {\bibfield  {journal} {\bibinfo  {journal}
  {Applied Physics Letters}\ }\textbf {\bibinfo {volume} {94}},\ \bibinfo
  {pages} {062107} (\bibinfo {year} {2009})}\BibitemShut {NoStop}%
\bibitem [{\citenamefont {Luxmoore}\ \emph {et~al.}(2014)\citenamefont
  {Luxmoore}, \citenamefont {Gan}, \citenamefont {Liu}, \citenamefont
  {Valmorra}, \citenamefont {Li}, \citenamefont {Faist},\ and\ \citenamefont
  {Nash}}]{luxmoore2014strong}%
  \BibitemOpen
  \bibfield  {author} {\bibinfo {author} {\bibfnamefont {I.~J.}\ \bibnamefont
  {Luxmoore}}, \bibinfo {author} {\bibfnamefont {C.~H.}\ \bibnamefont {Gan}},
  \bibinfo {author} {\bibfnamefont {P.~Q.}\ \bibnamefont {Liu}}, \bibinfo
  {author} {\bibfnamefont {F.}~\bibnamefont {Valmorra}}, \bibinfo {author}
  {\bibfnamefont {P.}~\bibnamefont {Li}}, \bibinfo {author} {\bibfnamefont
  {J.}~\bibnamefont {Faist}}, \ and\ \bibinfo {author} {\bibfnamefont {G.~R.}\
  \bibnamefont {Nash}},\ }\bibfield  {title} {\enquote {\bibinfo {title}
  {Strong coupling in the far-infrared between graphene plasmons and the
  surface optical phonons of silicon dioxide},}\ }\href@noop {} {\bibfield
  {journal} {\bibinfo  {journal} {ACS photonics}\ }\textbf {\bibinfo {volume}
  {1}},\ \bibinfo {pages} {1151--1155} (\bibinfo {year} {2014})}\BibitemShut
  {NoStop}%
\bibitem [{\citenamefont {Falkovsky}(2008)}]{falkovsky2008optical}%
  \BibitemOpen
  \bibfield  {author} {\bibinfo {author} {\bibfnamefont {L.}~\bibnamefont
  {Falkovsky}},\ }\bibfield  {title} {\enquote {\bibinfo {title} {Optical
  properties of graphene},}\ }in\ \href@noop {} {\emph {\bibinfo {booktitle}
  {Journal of Physics: Conference Series}}},\ Vol.\ \bibinfo {volume} {129}\
  (\bibinfo {organization} {IOP Publishing},\ \bibinfo {year} {2008})\ p.\
  \bibinfo {pages} {012004}\BibitemShut {NoStop}%
\bibitem [{\citenamefont {Papadakis}(2018)}]{papadakis2018optical}%
  \BibitemOpen
  \bibfield  {author} {\bibinfo {author} {\bibfnamefont {G.~T.}\ \bibnamefont
  {Papadakis}},\ }\emph {\bibinfo {title} {Optical Response in Planar
  Heterostructures: From Artificial Magnetism to Angstrom-Scale
  Metamaterials}},\ \href@noop {} {Ph.D. thesis},\ \bibinfo  {school}
  {California Institute of Technology} (\bibinfo {year} {2018})\BibitemShut
  {NoStop}%
\bibitem [{\citenamefont {Pochi~Yeh}(1988)}]{wiley1988sons}%
  \BibitemOpen
  \bibfield  {author} {\bibinfo {author} {\bibfnamefont {A.}~\bibnamefont
  {Pochi~Yeh}},\ }\href@noop {} {\enquote {\bibinfo {title} {Optical waves in
  layered media},}\ } (\bibinfo {year} {1988})\BibitemShut {NoStop}%
\bibitem [{\citenamefont {Nam}\ \emph {et~al.}(2012)\citenamefont {Nam},
  \citenamefont {Lindvall}, \citenamefont {Sun}, \citenamefont {Park},\ and\
  \citenamefont {Yurgens}}]{nam2012graphene}%
  \BibitemOpen
  \bibfield  {author} {\bibinfo {author} {\bibfnamefont {Y.}~\bibnamefont
  {Nam}}, \bibinfo {author} {\bibfnamefont {N.}~\bibnamefont {Lindvall}},
  \bibinfo {author} {\bibfnamefont {J.}~\bibnamefont {Sun}}, \bibinfo {author}
  {\bibfnamefont {Y.~W.}\ \bibnamefont {Park}}, \ and\ \bibinfo {author}
  {\bibfnamefont {A.}~\bibnamefont {Yurgens}},\ }\bibfield  {title} {\enquote
  {\bibinfo {title} {Graphene p--n--p junctions controlled by local gates made
  of naturally oxidized thin aluminium films},}\ }\href@noop {} {\bibfield
  {journal} {\bibinfo  {journal} {Carbon}\ }\textbf {\bibinfo {volume} {50}},\
  \bibinfo {pages} {1987--1992} (\bibinfo {year} {2012})}\BibitemShut {NoStop}%
\bibitem [{\citenamefont {Yusuf}\ \emph {et~al.}(2014)\citenamefont {Yusuf},
  \citenamefont {Nielsen}, \citenamefont {Dawber},\ and\ \citenamefont
  {Du}}]{yusuf2014extrinsic}%
  \BibitemOpen
  \bibfield  {author} {\bibinfo {author} {\bibfnamefont {M.~H.}\ \bibnamefont
  {Yusuf}}, \bibinfo {author} {\bibfnamefont {B.}~\bibnamefont {Nielsen}},
  \bibinfo {author} {\bibfnamefont {M.}~\bibnamefont {Dawber}}, \ and\ \bibinfo
  {author} {\bibfnamefont {X.}~\bibnamefont {Du}},\ }\bibfield  {title}
  {\enquote {\bibinfo {title} {Extrinsic and intrinsic charge trapping at the
  graphene/ferroelectric interface},}\ }\href@noop {} {\bibfield  {journal}
  {\bibinfo  {journal} {Nano letters}\ }\textbf {\bibinfo {volume} {14}},\
  \bibinfo {pages} {5437--5444} (\bibinfo {year} {2014})}\BibitemShut {NoStop}%
\bibitem [{\citenamefont {Pendharker}\ \emph {et~al.}(2017)\citenamefont
  {Pendharker}, \citenamefont {Hu}, \citenamefont {Molesky}, \citenamefont
  {Starko-Bowes}, \citenamefont {Poursoti}, \citenamefont {Pramanik},
  \citenamefont {Nazemifard}, \citenamefont {Fedosejevs}, \citenamefont
  {Thundat},\ and\ \citenamefont {Jacob}}]{pendharker2017thermal}%
  \BibitemOpen
  \bibfield  {author} {\bibinfo {author} {\bibfnamefont {S.}~\bibnamefont
  {Pendharker}}, \bibinfo {author} {\bibfnamefont {H.}~\bibnamefont {Hu}},
  \bibinfo {author} {\bibfnamefont {S.}~\bibnamefont {Molesky}}, \bibinfo
  {author} {\bibfnamefont {R.}~\bibnamefont {Starko-Bowes}}, \bibinfo {author}
  {\bibfnamefont {Z.}~\bibnamefont {Poursoti}}, \bibinfo {author}
  {\bibfnamefont {S.}~\bibnamefont {Pramanik}}, \bibinfo {author}
  {\bibfnamefont {N.}~\bibnamefont {Nazemifard}}, \bibinfo {author}
  {\bibfnamefont {R.}~\bibnamefont {Fedosejevs}}, \bibinfo {author}
  {\bibfnamefont {T.}~\bibnamefont {Thundat}}, \ and\ \bibinfo {author}
  {\bibfnamefont {Z.}~\bibnamefont {Jacob}},\ }\bibfield  {title} {\enquote
  {\bibinfo {title} {Thermal graphene metamaterials and epsilon-near-zero high
  temperature plasmonics},}\ }\href@noop {} {\bibfield  {journal} {\bibinfo
  {journal} {Journal of Optics}\ }\textbf {\bibinfo {volume} {19}},\ \bibinfo
  {pages} {055101} (\bibinfo {year} {2017})}\BibitemShut {NoStop}%
\bibitem [{\citenamefont {Dyachenko}\ \emph {et~al.}(2016)\citenamefont
  {Dyachenko}, \citenamefont {Molesky}, \citenamefont {Petrov}, \citenamefont
  {St{\"o}rmer}, \citenamefont {Krekeler}, \citenamefont {Lang}, \citenamefont
  {Ritter}, \citenamefont {Jacob},\ and\ \citenamefont
  {Eich}}]{dyachenko2016controlling}%
  \BibitemOpen
  \bibfield  {author} {\bibinfo {author} {\bibfnamefont {P.~N.}\ \bibnamefont
  {Dyachenko}}, \bibinfo {author} {\bibfnamefont {S.}~\bibnamefont {Molesky}},
  \bibinfo {author} {\bibfnamefont {A.~Y.}\ \bibnamefont {Petrov}}, \bibinfo
  {author} {\bibfnamefont {M.}~\bibnamefont {St{\"o}rmer}}, \bibinfo {author}
  {\bibfnamefont {T.}~\bibnamefont {Krekeler}}, \bibinfo {author}
  {\bibfnamefont {S.}~\bibnamefont {Lang}}, \bibinfo {author} {\bibfnamefont
  {M.}~\bibnamefont {Ritter}}, \bibinfo {author} {\bibfnamefont
  {Z.}~\bibnamefont {Jacob}}, \ and\ \bibinfo {author} {\bibfnamefont
  {M.}~\bibnamefont {Eich}},\ }\bibfield  {title} {\enquote {\bibinfo {title}
  {Controlling thermal emission with refractory epsilon-near-zero metamaterials
  via topological transitions},}\ }\href@noop {} {\bibfield  {journal}
  {\bibinfo  {journal} {Nature communications}\ }\textbf {\bibinfo {volume}
  {7}},\ \bibinfo {pages} {11809} (\bibinfo {year} {2016})}\BibitemShut
  {NoStop}%
\bibitem [{\citenamefont {Lobet}\ \emph {et~al.}(2016)\citenamefont {Lobet},
  \citenamefont {Majerus}, \citenamefont {Henrard},\ and\ \citenamefont
  {Lambin}}]{lobet2016perfect}%
  \BibitemOpen
  \bibfield  {author} {\bibinfo {author} {\bibfnamefont {M.}~\bibnamefont
  {Lobet}}, \bibinfo {author} {\bibfnamefont {B.}~\bibnamefont {Majerus}},
  \bibinfo {author} {\bibfnamefont {L.}~\bibnamefont {Henrard}}, \ and\
  \bibinfo {author} {\bibfnamefont {P.}~\bibnamefont {Lambin}},\ }\bibfield
  {title} {\enquote {\bibinfo {title} {Perfect electromagnetic absorption using
  graphene and epsilon-near-zero metamaterials},}\ }\href@noop {} {\bibfield
  {journal} {\bibinfo  {journal} {Physical Review B}\ }\textbf {\bibinfo
  {volume} {93}},\ \bibinfo {pages} {235424} (\bibinfo {year}
  {2016})}\BibitemShut {NoStop}%
\bibitem [{\citenamefont {Veselago}(1964)}]{veselago1964soviet}%
  \BibitemOpen
  \bibfield  {author} {\bibinfo {author} {\bibfnamefont {V.}~\bibnamefont
  {Veselago}},\ }\bibfield  {title} {\enquote {\bibinfo {title} {Soviet physics
  uspekhi, volume 10, number 4 january-february 1968.“},}\ }\href@noop {}
  {\bibfield  {journal} {\bibinfo  {journal} {The Electrodynamics Of Substances
  With Simultaneously Negative values Of E And $\mu$” PN Lebedev Physics
  Institute, Academy Of Sciences, USSR Usp. Fiz. Nauk}\ }\textbf {\bibinfo
  {volume} {92}},\ \bibinfo {pages} {517--526} (\bibinfo {year}
  {1964})}\BibitemShut {NoStop}%
\bibitem [{\citenamefont {Pendry}(2000)}]{pendry2000negative}%
  \BibitemOpen
  \bibfield  {author} {\bibinfo {author} {\bibfnamefont {J.~B.}\ \bibnamefont
  {Pendry}},\ }\bibfield  {title} {\enquote {\bibinfo {title} {Negative
  refraction makes a perfect lens},}\ }\href@noop {} {\bibfield  {journal}
  {\bibinfo  {journal} {Physical review letters}\ }\textbf {\bibinfo {volume}
  {85}},\ \bibinfo {pages} {3966} (\bibinfo {year} {2000})}\BibitemShut
  {NoStop}%
\bibitem [{\citenamefont {Choi}\ \emph {et~al.}(2011)\citenamefont {Choi},
  \citenamefont {Lee}, \citenamefont {Kim}, \citenamefont {Kang}, \citenamefont
  {Shin}, \citenamefont {Kwak}, \citenamefont {Kang}, \citenamefont {Lee},
  \citenamefont {Park},\ and\ \citenamefont {Min}}]{choi2011terahertz}%
  \BibitemOpen
  \bibfield  {author} {\bibinfo {author} {\bibfnamefont {M.}~\bibnamefont
  {Choi}}, \bibinfo {author} {\bibfnamefont {S.~H.}\ \bibnamefont {Lee}},
  \bibinfo {author} {\bibfnamefont {Y.}~\bibnamefont {Kim}}, \bibinfo {author}
  {\bibfnamefont {S.~B.}\ \bibnamefont {Kang}}, \bibinfo {author}
  {\bibfnamefont {J.}~\bibnamefont {Shin}}, \bibinfo {author} {\bibfnamefont
  {M.~H.}\ \bibnamefont {Kwak}}, \bibinfo {author} {\bibfnamefont {K.-Y.}\
  \bibnamefont {Kang}}, \bibinfo {author} {\bibfnamefont {Y.-H.}\ \bibnamefont
  {Lee}}, \bibinfo {author} {\bibfnamefont {N.}~\bibnamefont {Park}}, \ and\
  \bibinfo {author} {\bibfnamefont {B.}~\bibnamefont {Min}},\ }\bibfield
  {title} {\enquote {\bibinfo {title} {A terahertz metamaterial with
  unnaturally high refractive index},}\ }\href@noop {} {\bibfield  {journal}
  {\bibinfo  {journal} {Nature}\ }\textbf {\bibinfo {volume} {470}},\ \bibinfo
  {pages} {369} (\bibinfo {year} {2011})}\BibitemShut {NoStop}%
\bibitem [{\citenamefont {Drachev}, \citenamefont {Podolskiy},\ and\
  \citenamefont {Kildishev}(2013)}]{drachev2013hyperbolic}%
  \BibitemOpen
  \bibfield  {author} {\bibinfo {author} {\bibfnamefont {V.~P.}\ \bibnamefont
  {Drachev}}, \bibinfo {author} {\bibfnamefont {V.~A.}\ \bibnamefont
  {Podolskiy}}, \ and\ \bibinfo {author} {\bibfnamefont {A.~V.}\ \bibnamefont
  {Kildishev}},\ }\bibfield  {title} {\enquote {\bibinfo {title} {Hyperbolic
  metamaterials: new physics behind a classical problem},}\ }\href@noop {}
  {\bibfield  {journal} {\bibinfo  {journal} {Optics express}\ }\textbf
  {\bibinfo {volume} {21}},\ \bibinfo {pages} {15048--15064} (\bibinfo {year}
  {2013})}\BibitemShut {NoStop}%
\bibitem [{\citenamefont {Zhou}\ \emph {et~al.}(2005)\citenamefont {Zhou},
  \citenamefont {Koschny}, \citenamefont {Kafesaki}, \citenamefont {Economou},
  \citenamefont {Pendry},\ and\ \citenamefont
  {Soukoulis}}]{zhou2005saturation}%
  \BibitemOpen
  \bibfield  {author} {\bibinfo {author} {\bibfnamefont {J.}~\bibnamefont
  {Zhou}}, \bibinfo {author} {\bibfnamefont {T.}~\bibnamefont {Koschny}},
  \bibinfo {author} {\bibfnamefont {M.}~\bibnamefont {Kafesaki}}, \bibinfo
  {author} {\bibfnamefont {E.}~\bibnamefont {Economou}}, \bibinfo {author}
  {\bibfnamefont {J.}~\bibnamefont {Pendry}}, \ and\ \bibinfo {author}
  {\bibfnamefont {C.}~\bibnamefont {Soukoulis}},\ }\bibfield  {title} {\enquote
  {\bibinfo {title} {Saturation of the magnetic response of split-ring
  resonators at optical frequencies},}\ }\href@noop {} {\bibfield  {journal}
  {\bibinfo  {journal} {Physical review letters}\ }\textbf {\bibinfo {volume}
  {95}},\ \bibinfo {pages} {223902} (\bibinfo {year} {2005})}\BibitemShut
  {NoStop}%
\bibitem [{\citenamefont {Krishnamoorthy}\ \emph {et~al.}(2012)\citenamefont
  {Krishnamoorthy}, \citenamefont {Jacob}, \citenamefont {Narimanov},
  \citenamefont {Kretzschmar},\ and\ \citenamefont
  {Menon}}]{krishnamoorthy2012topological}%
  \BibitemOpen
  \bibfield  {author} {\bibinfo {author} {\bibfnamefont {H.~N.}\ \bibnamefont
  {Krishnamoorthy}}, \bibinfo {author} {\bibfnamefont {Z.}~\bibnamefont
  {Jacob}}, \bibinfo {author} {\bibfnamefont {E.}~\bibnamefont {Narimanov}},
  \bibinfo {author} {\bibfnamefont {I.}~\bibnamefont {Kretzschmar}}, \ and\
  \bibinfo {author} {\bibfnamefont {V.~M.}\ \bibnamefont {Menon}},\ }\bibfield
  {title} {\enquote {\bibinfo {title} {Topological transitions in
  metamaterials},}\ }\href@noop {} {\bibfield  {journal} {\bibinfo  {journal}
  {Science}\ }\textbf {\bibinfo {volume} {336}},\ \bibinfo {pages} {205--209}
  (\bibinfo {year} {2012})}\BibitemShut {NoStop}%
\bibitem [{\citenamefont {Gao}\ \emph {et~al.}(2015)\citenamefont {Gao},
  \citenamefont {Lawrence}, \citenamefont {Yang}, \citenamefont {Liu},
  \citenamefont {Fang}, \citenamefont {B{\'e}ri}, \citenamefont {Li},\ and\
  \citenamefont {Zhang}}]{gao2015topological}%
  \BibitemOpen
  \bibfield  {author} {\bibinfo {author} {\bibfnamefont {W.}~\bibnamefont
  {Gao}}, \bibinfo {author} {\bibfnamefont {M.}~\bibnamefont {Lawrence}},
  \bibinfo {author} {\bibfnamefont {B.}~\bibnamefont {Yang}}, \bibinfo {author}
  {\bibfnamefont {F.}~\bibnamefont {Liu}}, \bibinfo {author} {\bibfnamefont
  {F.}~\bibnamefont {Fang}}, \bibinfo {author} {\bibfnamefont {B.}~\bibnamefont
  {B{\'e}ri}}, \bibinfo {author} {\bibfnamefont {J.}~\bibnamefont {Li}}, \ and\
  \bibinfo {author} {\bibfnamefont {S.}~\bibnamefont {Zhang}},\ }\bibfield
  {title} {\enquote {\bibinfo {title} {Topological photonic phase in chiral
  hyperbolic metamaterials},}\ }\href@noop {} {\bibfield  {journal} {\bibinfo
  {journal} {Physical review letters}\ }\textbf {\bibinfo {volume} {114}},\
  \bibinfo {pages} {037402} (\bibinfo {year} {2015})}\BibitemShut {NoStop}%
\bibitem [{\citenamefont {Tay}\ \emph {et~al.}(2008)\citenamefont {Tay},
  \citenamefont {Blanche}, \citenamefont {Voorakaranam}, \citenamefont
  {Tun{\c{c}}}, \citenamefont {Lin}, \citenamefont {Rokutanda}, \citenamefont
  {Gu}, \citenamefont {Flores}, \citenamefont {Wang}, \citenamefont {Li} \emph
  {et~al.}}]{tay2008updatable}%
  \BibitemOpen
  \bibfield  {author} {\bibinfo {author} {\bibfnamefont {S.}~\bibnamefont
  {Tay}}, \bibinfo {author} {\bibfnamefont {P.-A.}\ \bibnamefont {Blanche}},
  \bibinfo {author} {\bibfnamefont {R.}~\bibnamefont {Voorakaranam}}, \bibinfo
  {author} {\bibfnamefont {A.}~\bibnamefont {Tun{\c{c}}}}, \bibinfo {author}
  {\bibfnamefont {W.}~\bibnamefont {Lin}}, \bibinfo {author} {\bibfnamefont
  {S.}~\bibnamefont {Rokutanda}}, \bibinfo {author} {\bibfnamefont
  {T.}~\bibnamefont {Gu}}, \bibinfo {author} {\bibfnamefont {D.}~\bibnamefont
  {Flores}}, \bibinfo {author} {\bibfnamefont {P.}~\bibnamefont {Wang}},
  \bibinfo {author} {\bibfnamefont {G.}~\bibnamefont {Li}},  \emph {et~al.},\
  }\bibfield  {title} {\enquote {\bibinfo {title} {An updatable holographic
  three-dimensional display},}\ }\href@noop {} {\bibfield  {journal} {\bibinfo
  {journal} {Nature}\ }\textbf {\bibinfo {volume} {451}},\ \bibinfo {pages}
  {694} (\bibinfo {year} {2008})}\BibitemShut {NoStop}%
\bibitem [{\citenamefont {Lutkenhaus}\ \emph {et~al.}(2013)\citenamefont
  {Lutkenhaus}, \citenamefont {George}, \citenamefont {Moazzezi}, \citenamefont
  {Philipose},\ and\ \citenamefont {Lin}}]{lutkenhaus2013digitally}%
  \BibitemOpen
  \bibfield  {author} {\bibinfo {author} {\bibfnamefont {J.}~\bibnamefont
  {Lutkenhaus}}, \bibinfo {author} {\bibfnamefont {D.}~\bibnamefont {George}},
  \bibinfo {author} {\bibfnamefont {M.}~\bibnamefont {Moazzezi}}, \bibinfo
  {author} {\bibfnamefont {U.}~\bibnamefont {Philipose}}, \ and\ \bibinfo
  {author} {\bibfnamefont {Y.}~\bibnamefont {Lin}},\ }\bibfield  {title}
  {\enquote {\bibinfo {title} {Digitally tunable holographic lithography using
  a spatial light modulator as a programmable phase mask},}\ }\href@noop {}
  {\bibfield  {journal} {\bibinfo  {journal} {Optics express}\ }\textbf
  {\bibinfo {volume} {21}},\ \bibinfo {pages} {26227--26235} (\bibinfo {year}
  {2013})}\BibitemShut {NoStop}%
\bibitem [{\citenamefont {Khan}\ and\ \citenamefont
  {Johnson}(2014)}]{khan2014lifshitz}%
  \BibitemOpen
  \bibfield  {author} {\bibinfo {author} {\bibfnamefont {S.~N.}\ \bibnamefont
  {Khan}}\ and\ \bibinfo {author} {\bibfnamefont {D.~D.}\ \bibnamefont
  {Johnson}},\ }\bibfield  {title} {\enquote {\bibinfo {title} {Lifshitz
  transition and chemical instabilities in ba 1- x k x fe 2 as 2
  superconductors},}\ }\href@noop {} {\bibfield  {journal} {\bibinfo  {journal}
  {Physical review letters}\ }\textbf {\bibinfo {volume} {112}},\ \bibinfo
  {pages} {156401} (\bibinfo {year} {2014})}\BibitemShut {NoStop}%
\bibitem [{\citenamefont {Rechtsman}\ \emph {et~al.}(2013)\citenamefont
  {Rechtsman}, \citenamefont {Zeuner}, \citenamefont {Plotnik}, \citenamefont
  {Lumer}, \citenamefont {Podolsky}, \citenamefont {Dreisow}, \citenamefont
  {Nolte}, \citenamefont {Segev},\ and\ \citenamefont
  {Szameit}}]{rechtsman2013photonic}%
  \BibitemOpen
  \bibfield  {author} {\bibinfo {author} {\bibfnamefont {M.~C.}\ \bibnamefont
  {Rechtsman}}, \bibinfo {author} {\bibfnamefont {J.~M.}\ \bibnamefont
  {Zeuner}}, \bibinfo {author} {\bibfnamefont {Y.}~\bibnamefont {Plotnik}},
  \bibinfo {author} {\bibfnamefont {Y.}~\bibnamefont {Lumer}}, \bibinfo
  {author} {\bibfnamefont {D.}~\bibnamefont {Podolsky}}, \bibinfo {author}
  {\bibfnamefont {F.}~\bibnamefont {Dreisow}}, \bibinfo {author} {\bibfnamefont
  {S.}~\bibnamefont {Nolte}}, \bibinfo {author} {\bibfnamefont
  {M.}~\bibnamefont {Segev}}, \ and\ \bibinfo {author} {\bibfnamefont
  {A.}~\bibnamefont {Szameit}},\ }\bibfield  {title} {\enquote {\bibinfo
  {title} {Photonic floquet topological insulators},}\ }\href@noop {}
  {\bibfield  {journal} {\bibinfo  {journal} {Nature}\ }\textbf {\bibinfo
  {volume} {496}},\ \bibinfo {pages} {196} (\bibinfo {year}
  {2013})}\BibitemShut {NoStop}%
\bibitem [{\citenamefont {Li}\ and\ \citenamefont {Mei}(2015)}]{li2015double}%
  \BibitemOpen
  \bibfield  {author} {\bibinfo {author} {\bibfnamefont {Y.}~\bibnamefont
  {Li}}\ and\ \bibinfo {author} {\bibfnamefont {J.}~\bibnamefont {Mei}},\
  }\bibfield  {title} {\enquote {\bibinfo {title} {Double dirac cones in
  two-dimensional dielectric photonic crystals},}\ }\href@noop {} {\bibfield
  {journal} {\bibinfo  {journal} {Optics express}\ }\textbf {\bibinfo {volume}
  {23}},\ \bibinfo {pages} {12089--12099} (\bibinfo {year} {2015})}\BibitemShut
  {NoStop}%
\bibitem [{\citenamefont {Huang}\ \emph {et~al.}(2011)\citenamefont {Huang},
  \citenamefont {Lai}, \citenamefont {Hang}, \citenamefont {Zheng},\ and\
  \citenamefont {Chan}}]{huang2011dirac}%
  \BibitemOpen
  \bibfield  {author} {\bibinfo {author} {\bibfnamefont {X.}~\bibnamefont
  {Huang}}, \bibinfo {author} {\bibfnamefont {Y.}~\bibnamefont {Lai}}, \bibinfo
  {author} {\bibfnamefont {Z.~H.}\ \bibnamefont {Hang}}, \bibinfo {author}
  {\bibfnamefont {H.}~\bibnamefont {Zheng}}, \ and\ \bibinfo {author}
  {\bibfnamefont {C.}~\bibnamefont {Chan}},\ }\bibfield  {title} {\enquote
  {\bibinfo {title} {Dirac cones induced by accidental degeneracy in photonic
  crystals and zero-refractive-index materials},}\ }\href@noop {} {\bibfield
  {journal} {\bibinfo  {journal} {Nature materials}\ }\textbf {\bibinfo
  {volume} {10}},\ \bibinfo {pages} {582} (\bibinfo {year} {2011})}\BibitemShut
  {NoStop}%
\bibitem [{\citenamefont {Dai}\ \emph {et~al.}(2015{\natexlab{b}})\citenamefont
  {Dai}, \citenamefont {Ma}, \citenamefont {Liu}, \citenamefont {Andersen},
  \citenamefont {Fei}, \citenamefont {Goldflam}, \citenamefont {Wagner},
  \citenamefont {Watanabe}, \citenamefont {Taniguchi}, \citenamefont {Thiemens}
  \emph {et~al.}}]{dai2015graphene}%
  \BibitemOpen
  \bibfield  {author} {\bibinfo {author} {\bibfnamefont {S.}~\bibnamefont
  {Dai}}, \bibinfo {author} {\bibfnamefont {Q.}~\bibnamefont {Ma}}, \bibinfo
  {author} {\bibfnamefont {M.}~\bibnamefont {Liu}}, \bibinfo {author}
  {\bibfnamefont {T.}~\bibnamefont {Andersen}}, \bibinfo {author}
  {\bibfnamefont {Z.}~\bibnamefont {Fei}}, \bibinfo {author} {\bibfnamefont
  {M.}~\bibnamefont {Goldflam}}, \bibinfo {author} {\bibfnamefont
  {M.}~\bibnamefont {Wagner}}, \bibinfo {author} {\bibfnamefont
  {K.}~\bibnamefont {Watanabe}}, \bibinfo {author} {\bibfnamefont
  {T.}~\bibnamefont {Taniguchi}}, \bibinfo {author} {\bibfnamefont
  {M.}~\bibnamefont {Thiemens}},  \emph {et~al.},\ }\bibfield  {title}
  {\enquote {\bibinfo {title} {Graphene on hexagonal boron nitride as a tunable
  hyperbolic metamaterial},}\ }\href@noop {} {\bibfield  {journal} {\bibinfo
  {journal} {Nature nanotechnology}\ }\textbf {\bibinfo {volume} {10}},\
  \bibinfo {pages} {682} (\bibinfo {year} {2015}{\natexlab{b}})}\BibitemShut
  {NoStop}%
\end{thebibliography}

%

\end{document}